\documentclass[preprint,12pt,times]{elsarticle}
\journal{{\textrm Journal of Computational Physics}}

\usepackage{amssymb}
\usepackage{amsmath}
\usepackage{amsfonts}
\usepackage{bm}
\usepackage{bbm}
\usepackage{mathrsfs}
\usepackage{geometry}
\usepackage{xspace}
\usepackage{bm}
\usepackage{srcltx}
\usepackage{caption}
\usepackage[labelformat=simple]{subcaption}
\usepackage[usenames,dvipsnames,svgnames,table]{xcolor}
\usepackage[colorlinks,linkcolor=blue,citecolor=blue,urlcolor=blue]{hyperref} 
\renewcommand{\thetable}{\arabic{table}}
\usepackage{psfrag}
\usepackage{graphicx}

\usepackage{booktabs}
\usepackage{pgfplotstable}
\pgfplotsset{compat=newest}
\pgfplotstableset{
    fixed zerofill, precision=4,
    every head row/.style={before row=\toprule,after row=\midrule},
    every last row/.style={after row=\bottomrule}
}

\usepackage[lined,commentsnumbered]{algorithm2e}
\usepackage[inline]{enumitem}
\usepackage{nicefrac}

\biboptions{sort&compress,comma,square}

\usepackage[figuresright]{rotating}

\renewcommand\thesubfigure{(\alph{subfigure})}
\def\gz  #1{           \mbox{$\mathbf{#1}$}}
\DeclareMathAlphabet{\bsf}{OT1}{cmss}{bx}{n}

\def\Grad #1 {{\rm Grad} #1}
\def\grad #1 {{\rm grad} #1}

\def\d {\,\mbox{d}}

\newcommand{\norm}[1]{\left\lvert #1 \right\rvert}

\newcommand{\dbracket}[1]{\left[\!\!\left[ #1 \right]\!\!\right]}

\def\mcl  #1{               {\cal #1}}

\begin{document}
\begin{frontmatter}
\title{On the $h$\,-adaptive PUM and the $hp$\,-adaptive FEM approaches applied to PDEs in quantum mechanics}

\author[a]{Denis Davydov\corref{cor}}
\ead{denis.davydov@fau.de}

\author[b]{Tymofiy Gerasimov}
\ead{t.gerasimov@tu-braunschweig.de}

\author[a]{Jean-Paul Pelteret}
\ead{jean-paul.pelteret@ltm.uni-erlangen.de}

\author[a]{Paul Steinmann}
\ead{paul.steinmann@ltm.uni-erlangen.de}

\cortext[cor]{Corresponding author.}

\address[a]{Chair of Applied Mechanics, University of
    Erlangen-Nuremberg, Egerlandstr.\ 5, 91058 Erlangen, Germany}

\address[b]{Institute of Applied Mechanics,
    Technische Universit{\"a}t Braunschweig,
    Bienroder Weg 87, 38106 Braunschweig, Germany}

\begin{abstract}
In this paper the $h$\,-adaptive partition-of-unity method and the $h$\,- and $hp$\,-adaptive finite element method are applied to partial differential equations arising in quantum mechanics, 
namely, the Schr{\"o}dinger equation with Coulomb and harmonic potentials, and the Poisson problem.
Implementational details of the partition-of-unity method related to enforcing continuity with hanging nodes and the degeneracy of the basis are discussed.
The partition-of-unity method is equipped with an \textit{a posteriori} error estimator, thus enabling implementation of error-controlled \textit{adaptive} mesh refinement strategies.
To that end, local interpolation error estimates are derived for the partition-of-unity method enriched with a class of exponential functions. 
The results are the same as for the finite element method and thereby admit the usage of standard residual error indicators.
The efficiency of the $h$\,-adaptive partition-of-unity method is compared to the $h$\,- and $hp$\,-adaptive finite element method.
The latter is implemented by adopting the analyticity estimate from Legendre coefficients. 
An extension of this approach to multiple solution vectors is proposed.
Numerical results confirm the remarkable accuracy of the $h$\,-adaptive partition-of-unity approach.
In case of the Hydrogen atom, the $h$\,-adaptive linear partition-of-unity method was found to be comparable to the $hp$\,-adaptive finite element method for the target eigenvalue accuracy of $10^{-3}$. 

\end{abstract}


\end{frontmatter}

\section{Introduction}
\label{sec:intro}
Recently there has been an increase of interest in applying Finite Element (FE) methods to partial differential equations (PDEs) in quantum mechanics \cite{Maday2014,Davydov2014,Linder:2012tg,Motamarri:2012wm,Fang:2012uu,Pask:2005bf,Zhang:2008dp,Bylaska:2009im,Bao:2012it,Sukumar:2009vw,Fattebert:2007fy,White1989, Cimrman2015}, 
namely to the coupled eigenvalue and Poisson problems.
In order to improve the accuracy of the solution, 
the basis set can be adaptively expanded through either 
refinement of the mesh ($h$\,-adaptivity) 
or the basis functions can be
augmented by the introduction of higher polynomial degree basis
functions ($p$\,-adaptivity). 
Since the solution is not smooth and contains cusp singularities,
the application of the $h$\,-adaptive FEM may require very fine meshes and 
could be computationally inefficient.
There are several approaches to circumvent this problem.

From the physical point of view, for \textit{ab initio} calculation of molecules often core electrons (as opposed to valence electrons) behave in a similar way to single atom solutions. 
Thus one possesses an \textit{a priori} knowledge of a part of the solution vectors to the eigenvalue problem.
One of the approaches used to introduce this into a FE formulation is the Partition-of-Unity Method (PUM) \cite{Melenk1996,Babuska1997}, which is a generalization of the classical FE method.
In PUM the enrichment functions are introduced into a basis as products with standard FE shape functions, thereby enlarging the standard FE space.
As the standard FE functions satisfy the partition-of-unity property (that is, they sum to one in the whole domain), the resulting basis can reproduce enrichment functions exactly.
In the continuum mechanics community this method is known as XFEM \cite{Laborde2005,Fries2008,Chahine2008,Xiao2006,Gerasimov2012,Belytschko1999,Dolbow1999,Patzak2003}, originally popularized by Belytschko and Black \cite{Belytschko1999}. 
For an overview on this topic we refer the reader to \cite{Simone2007,Belytschko2009,Fries2010}.

An alternative approach to the above is to combine $h$\,- and $p$\,-adaptivity
resulting in what is termed as $hp$\,-adaptive FEM. 
For an overview of $hp$\,-adaptive refinement strategies we refer the reader to \cite{Mitchell2014}.
The general idea is that when the exact solution is smooth on the given element, $p$-adaptive refinement is more efficient and leads
to a faster convergence;  
whereas if the solution is non-smooth (singular), $h$\,-adaptive refinement is performed.
Thus in addition to a reliable error estimate and the choice of the marking strategy of elements for refinement,
$hp$\,-adaptive methods need to decide which type of refinement to perform on a given element.
In this work we use methods based on smoothness estimation \cite{Houston2005, Hartmann2010, Mavriplis1994, Eibner2007,Fankhauser2014,Bangerth:2008uw}.
As those methods are normally employed for problems with a single solution vector, 
we propose an extension to multiple solution vectors 
as is required for the here considered eigenvalue problems.

Herein, our main focus is application of $h$\,-adaptive PUM and $hp$\,-adaptive FEM to PDEs in quantum mechanics, namely to the Schr{\"o}dinger equation and the Poisson problem, and comparison of efficiency of these approaches.
Application of the PUM to the above problems holds a significant promise to improve on accuracy of a standard (non-enriched) FE approximation. 
The corresponding numerical evidence can be found in \cite{Pask2011,Sukumar:2009vw}, where convergence studies for PUM solutions obtained on \textit{uniformly} refined meshes are performed. 

In our paper, the PUM will be equipped with an \textit{a posteriori} error estimator, thus enabling implementation of error-controlled \textit{adaptive} mesh refinement strategies.
As for the model problems, 
we limit ourselves to uncoupled eigenvalue and Poisson problems as analytic solutions are available for that case. 
All findings are expected to apply to more complicated cases when the two equations are coupled,
such as those arising from the Density Functional Theory \cite{Hohenberg:1964ut,Kohn:1965ui}.

The outline of this paper is as follows: In section \ref{sec:problem} the considered PDEs and their solution are introduced. 
The PUM and its implementational details are given in Section \ref{sec:POU}.
Section \ref{sec:hp} is devoted to the strategy to decide between $h$\,- and $p$\,- adaptive refinement. 
Results of numerical studies of the chosen systems are 
presented in section \ref{sec:results}, followed by some conclusions in Section \ref{sec:summary}. 
Finally, in the Appendix we rigorously derive the local interpolation error estimates for enrichment with a class of exponential functions.

\section{Problem formulation}
\label{sec:problem}
In order to motivate the use of the PUM, it is necessary to understand some of the difficulties arising from the classes of problems that we will evaluate in this work.
In this manuscript we consider the following three-dimensional problems that have analytical solutions:
\subsection{Eigenvalue problem}
The eigenvalue problem that we will consider is the Schr{\"o}dinger equation, for which we seek lowest eigenpairs $(\lambda_\alpha,\psi_\alpha)$ of
\begin{align}
\begin{split}
\Bigg[-\frac{1}{2}  \nabla^2 + V(\gz x)  \Bigg]\, \psi_\alpha(\gz x) &= \lambda_\alpha \psi_\alpha(\gz x) \quad \rm{on}\; \Omega\,, \\
\psi_\alpha(\gz x) &= 0 \quad \rm{on}\; \partial \Omega,
\end{split}
\label{eq:Schroedinger}
\end{align}
with two different (spherical) potentials $V(\gz x)=V(\norm{\gz x})$\footnote{For spherically symmetric potentials one can separate eigenfunctions into radial $R_{n,l}(r)$ and angular $Y_{m,l}(\theta,\phi)$ parts, where the latter are spherical harmonics \cite{Griffiths:2005tq}. Here $\{n,l,m\}$ are three quantum numbers.
    }. 

The first case is the the Coulomb potential 
$V(\gz x) = -1/\norm{\gz x}$, which 
 corresponds to the Hydrogen atom. The eigenvalues of this problem are degenerate. In $\mathbb{R}^3$, on each energy level $n$ there are $n^2$ eigenvalues $\lambda_n = \lambda_1/n^2$, where 
 $\lambda_1 = -1/2$ 
  \cite{Griffiths:2005tq}. The eigenvector corresponding to the lowest eigenvalue reads
\begin{align}
\psi_1(\gz x) = \frac{1}{\sqrt{\pi}} \exp \left(-\norm{\gz x} \right)\;.
\label{eq:Hydrogenic1}
\end{align}
The 
radial component of the eigenfunctions at the next energy level are $R_{2,0}=[1-\norm{\gz x}/2]\exp(-\norm{\gz x}/2)$ and $R_{2,1}=\norm{\gz x}/2\exp(-\norm{\gz x}/2)$.

\begin{figure}[!ht]
    \begin{subfigure}[b]{0.49\textwidth}
        \centering
        \psfrag{x}[c][c]{$\norm{\gz x}$}
        \psfrag{psi}[c][c]{$\psi$}
        \psfrag{R10}[c][c][0.7]{$R_{1,0}$}
        \psfrag{R20}[c][c][0.7]{$R_{2,0}$}
        \psfrag{R21}[c][c][0.7]{$R_{2,1}$}
        \includegraphics[width=0.98\textwidth]{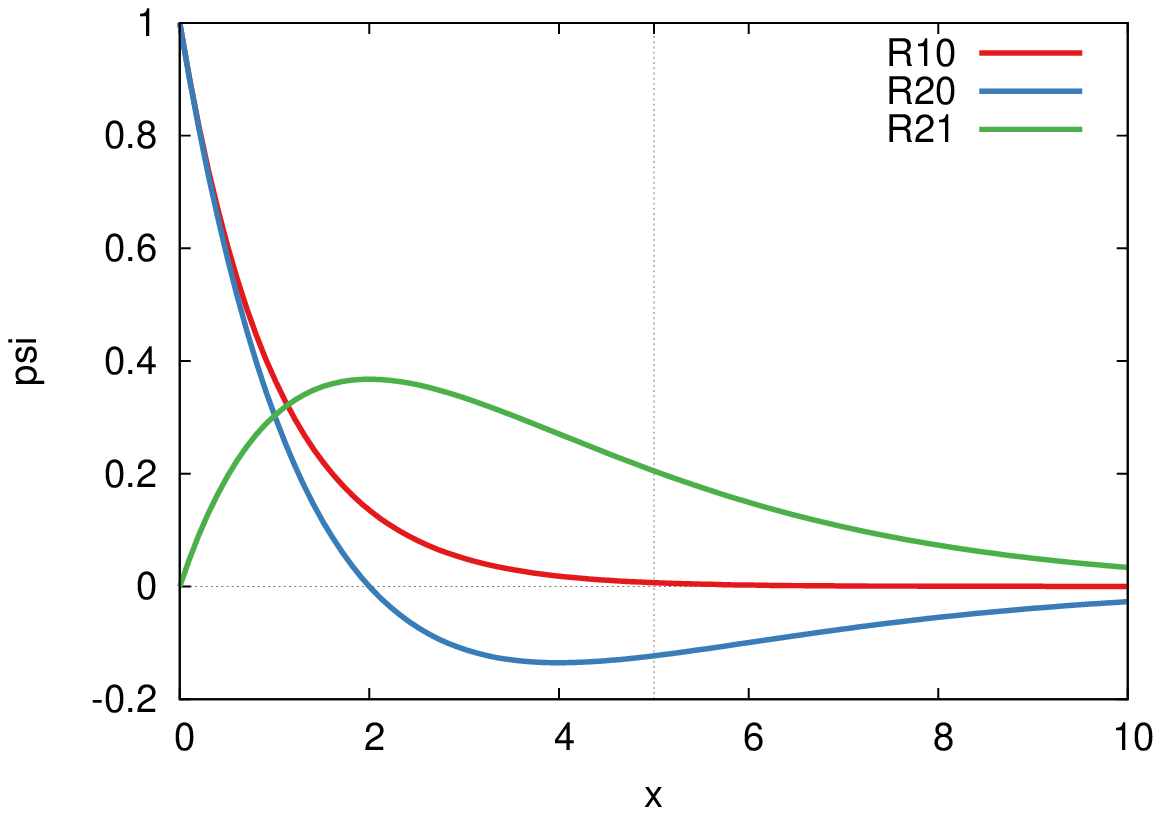}
        \caption{Coulomb}
        \label{fig:schroedinger_solution_C}
    \end{subfigure}  
    ~
    \begin{subfigure}[b]{0.49\textwidth}
        \centering
        \psfrag{x}[c][c]{$\norm{\gz x}$}
        \psfrag{psi}[c][c]{$\psi$}
        \psfrag{RH00}[c][c][0.7]{$R_{0,0}$}
        \psfrag{RH01}[c][c][0.7]{$R_{0,1}$}
        \includegraphics[width=0.98\textwidth]{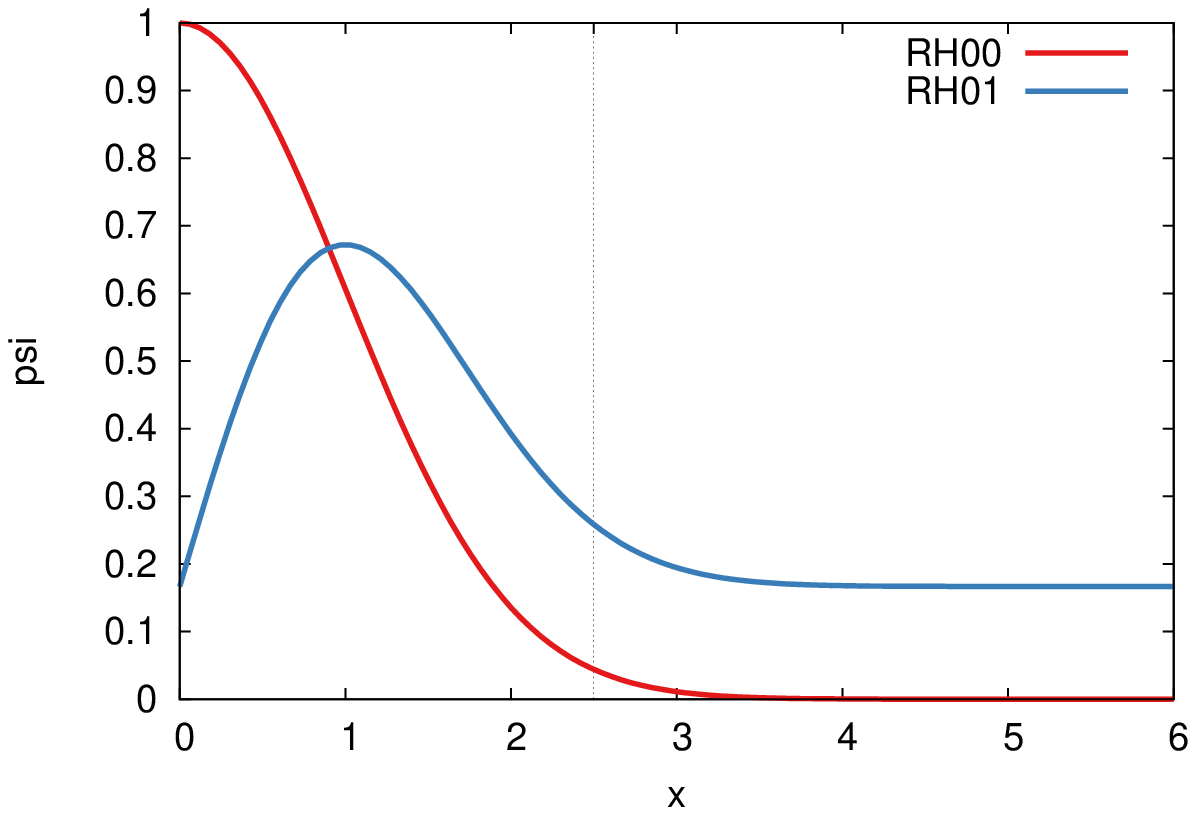}
        \caption{Harmonic}
        \label{fig:schroedinger_solution_H}
    \end{subfigure}
    \caption{Radial components of eigenfunctions for different potentials $V(\gz x)$.
        The dotted vertical line indicates the smallest initial mesh size which will be used in our numerical calculations.
    }
    \label{fig:schroedinger_solution}
\end{figure}

The second potential we will consider is a harmonic potential $V(\gz x)=\norm{\gz x}^2/2$ that leads to a harmonic oscillator problem. The eigenvalues for this problem are also degenerate; in $\mathbb{R}^3$ they are given by $\lambda_n = n+1/2$ for $n$-th energy level. The lowest two have a degeneracy of 1 and 3, respectively. The (unnormalized) eigenvector corresponding to the lowest eigenvalue is
\begin{align}
\psi_1(\gz x) = \exp \left(-\norm{\gz x}^2/2 \right)\;.
\label{eq:Harmonic1}
\end{align}
The radial component of the next eigenfunction is $R_{0,1}(\gz x) = \norm{\gz x} \exp \left(-\norm{\gz x}^2/2 \right)$.
Figure \ref{fig:schroedinger_solution} shows radial components of eigenfunctions for the Coulomb and harmonic potential.
It is clear that in order to have a low interpolation error for a standard Lagrange FE basis, a very fine mesh will be required near the origin.
For such non-smooth solutions we will see that by introducing enrichment functions the interpolation error of the resulting FE basis will be greatly reduced.

\subsection{Poisson problem}
\begin{figure}[!ht]
    \begin{subfigure}[b]{0.49\textwidth}
        \centering
        \psfrag{x}[c][c]{$\norm{\gz x}$}
        \psfrag{psi}[c][c]{$\phi$}
        \psfrag{rho}[c][c]{$\rho$}
        \includegraphics[width=0.98\textwidth]{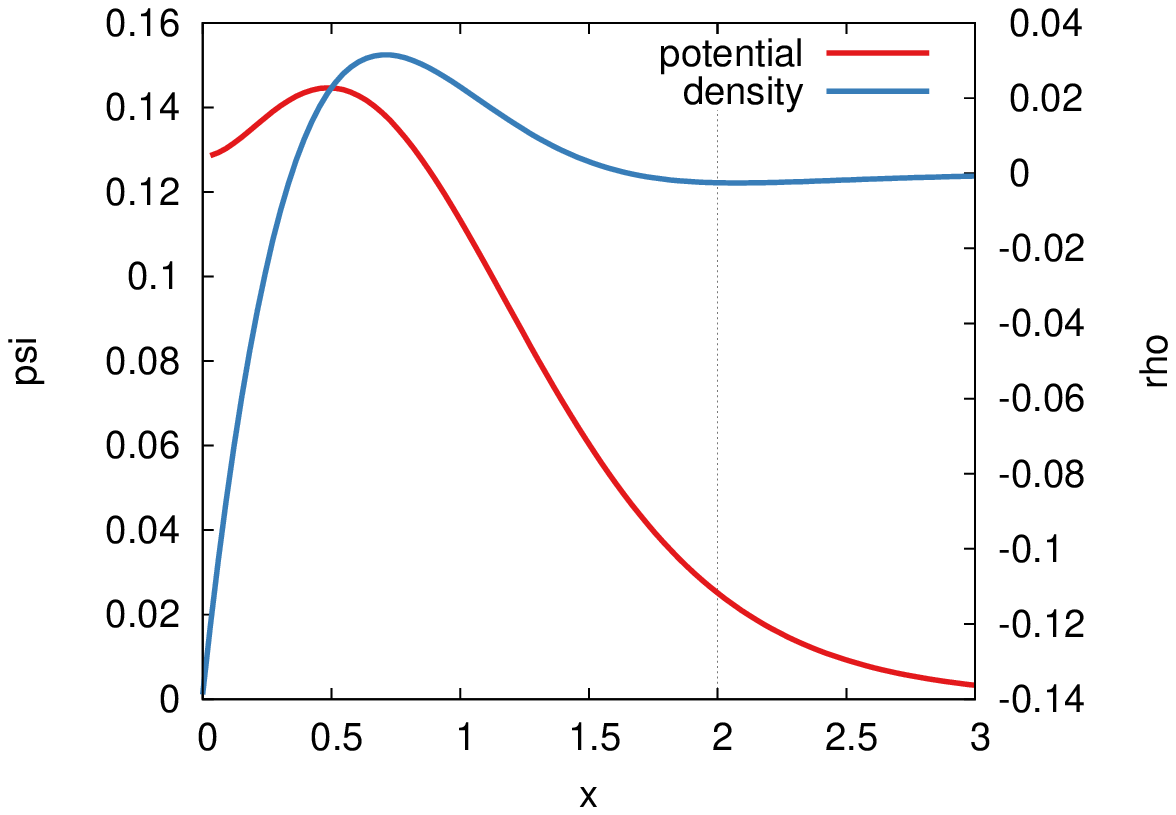}
        \caption{$\sigma=1.0$}
        \label{fig:density_solution10}
    \end{subfigure}  
    ~
    \begin{subfigure}[b]{0.49\textwidth}
        \centering
        \psfrag{x}[c][c]{$\norm{\gz x}$}
        \psfrag{psi}[c][c]{$\phi$}
        \psfrag{rho}[c][c]{$\rho$}
        \includegraphics[width=0.98\textwidth]{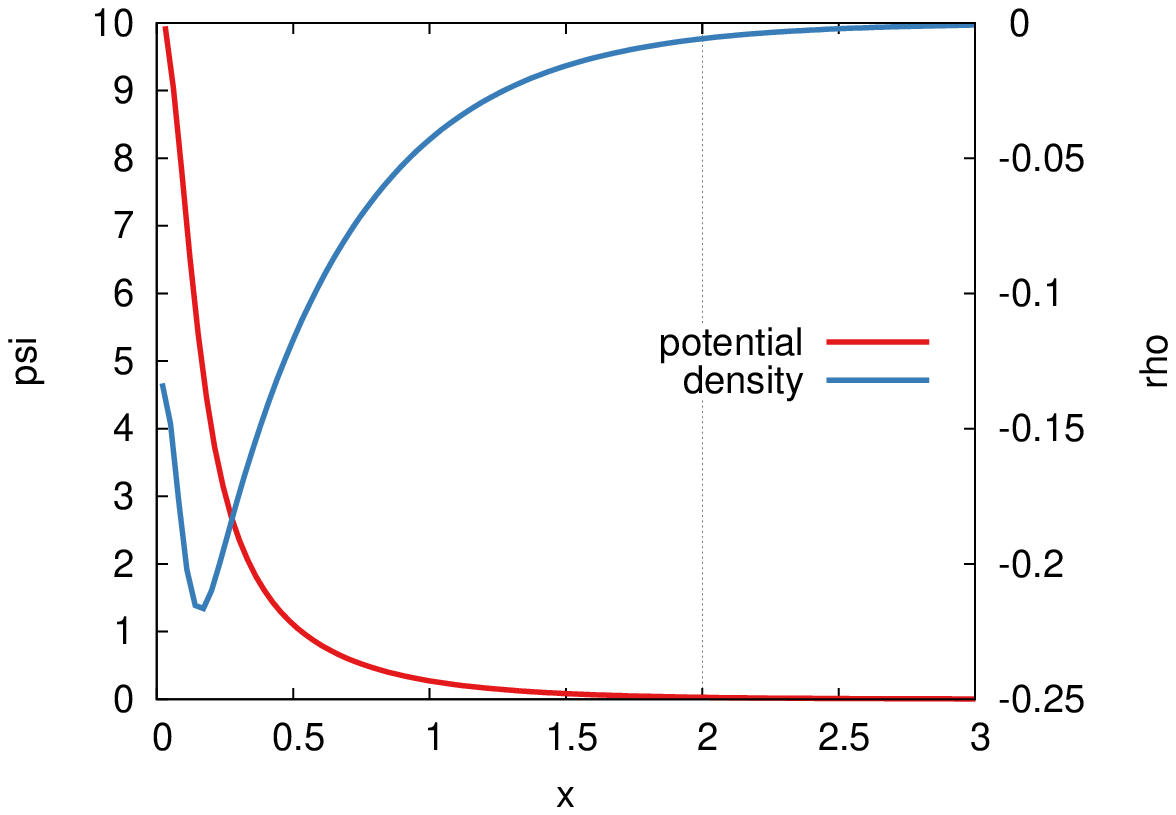}
        \caption{$\sigma=0.1$}
        \label{fig:density_solution01}
    \end{subfigure}
    \caption{Charge density and electrostatic potential for different values of $\sigma$.
        The dotted vertical line indicates the smallest initial mesh size which will be used in our numerical calculations.
    }
    \label{fig:density_solution}
\end{figure}

The associated Poisson equation relates the electron density field and the electrostatic potential. In atomic units it reads
\begin{align}
- \nabla^2 \phi(\gz x) = 4\pi \rho(\gz x) \quad \rm{on}\; \Omega\,.
\label{eq:Poisson}
\end{align}
The density function on the right-hand-side is composed of the squares of eigenvectors, possibly with the addition of other terms.
In case of the Hydrogen atom, the total charge density is composed of the electron density less the singular nucleus density
\begin{align}
\rho(\gz x) \equiv \psi_1^2 -\delta (\gz x)\;,
\end{align}
where $\psi_1$ is the electron wave-function given in (\ref{eq:Hydrogenic1}).
Note that $\rho(\gz x)$ is not in $H^{-1}$ and thus the solution $\phi(\gz x)$ is not in $H^1$.
The corresponding electrostatic potential produced reads
\begin{align}
\phi = -\exp\left(-2\norm{\gz x}\right)\left[1+\frac{1}{\norm{\gz x}}\right]\;.
\label{eq:poisson_sol_singular}
\end{align}
For the numerical analysis below we will consider a regularized counterpart where the delta function is substituted by a Gaussian distribution
\begin{align}
\rho(\gz x) \equiv \psi_1^2 -\frac{1}{\pi^{\nicefrac{3}{2}}\sigma^3}\exp\left( -\frac{\norm{\gz x}^2}{\sigma^2} \right) \;.
\end{align}
This corresponds to a split of the nuclei Coulomb potential into an (almost) local short range part and smooth long range part \cite{Davydov2014}. 
The electrostatic potential produced in this case reads
\begin{align}
\phi = -\exp\left(-2\norm{\gz x}\right)\left[1+\frac{1}{\norm{\gz x}}\right]+\frac{1}{\norm{\gz x}} - \frac{\rm{erf}(\norm{\gz x}/\sigma)}{\norm{\gz x}}\;.
\end{align}
Figure \ref{fig:density_solution} shows radial components of density and potential fields for different values of $\sigma$.
It is clear that by varying $\sigma$, the character of the solution is changed from smooth to more singular.
The limit $\sigma\rightarrow 0$ corresponds to the singular solution in Equation \ref{eq:poisson_sol_singular}.

\section{partition-of-unity method}
\label{sec:POU}
\subsection{Enriched FE space}
The classical FEM may fail when the solution is not smooth or is highly oscillatory. In either case, in order to obtain an accurate solution using piecewise polynomial spaces one has to employ a very fine mesh that increases the computational cost of solving the problem.
The PUM proposed by Melenk and Babuska in \cite{Melenk1996,Babuska1997} can address this issue. The main features of the PUM are (i) the inclusion of an \textit{a priori} knowledge about the solution into the FE space, and (ii) the construction of an FE space of any desired regularity. It is the former attribute which is important in the context of this work. 
The PUM enriches the vector space spanned by standard FE basis functions $N_i(\gz x)$ (e.g. polynomials) by products of these functions with functions $f_{j}(\gz x)$ that contain \textit{a-priori} knowledge about the solution
\begin{align}
u(\gz x) = \sum_{i \in I} N_i(\gz x) \left[ u_i + \sum_{j \in S} f_j(\gz x) \widetilde{u}_{ij} \right] .
\label{eq:POU}
\end{align}
Here $u_i$ are standard degrees-of-freedom (DoFs) and $\widetilde{u}_{ij}$ are additional DoFs associated with the shape functions $N_i(\gz x)$ and the enrichment functions $f_j(\gz x)$; $I$ is a set of all nodes and $S$ is the set of enrichment functions.
Since (possibly global) enrichment functions $f_j(\gz x)$ are multiplied with $N_i(\gz x)$ which has local support, the product also has local support and therefore matrices arising from the weak form remain sparse. Also, since the standard shape functions satisfy the partition of unity property $\sum_i N_i(\gz x) \equiv 1\;\rm{on}\;\Omega$, the resulting vector space can reproduce enrichment functions $f_j(\gz x)$ exactly.

\subsection{Implementational details}
\begin{figure}[!ht]
    \begin{subfigure}[b]{0.49\textwidth}
        \centering
        \includegraphics[width=0.98\textwidth]{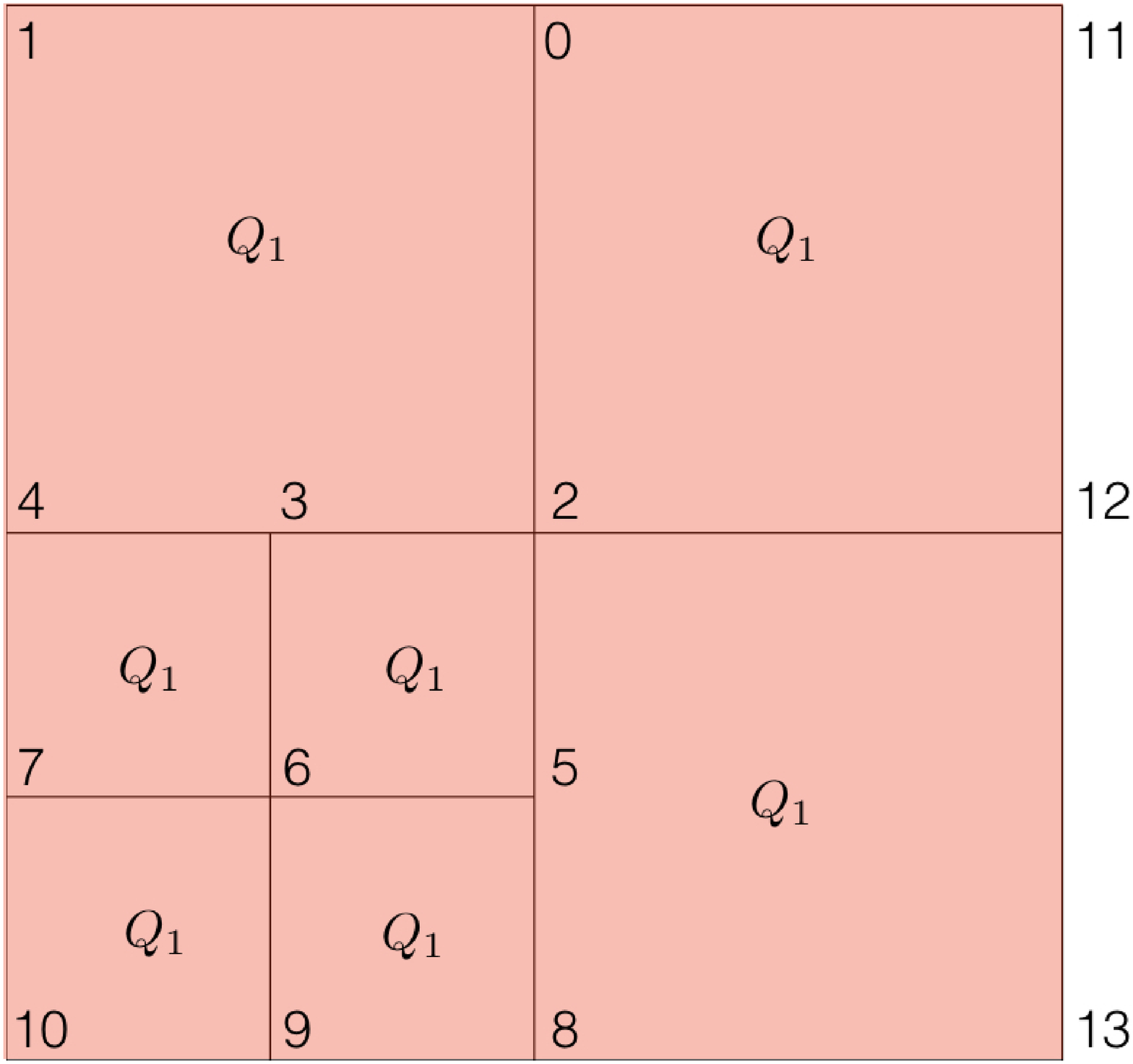}
        \caption{first FE space (standard)}
        \label{fig:pou_hanging_a}
    \end{subfigure}  
    ~
    \begin{subfigure}[b]{0.49\textwidth}
        \centering
        \includegraphics[width=0.98\textwidth]{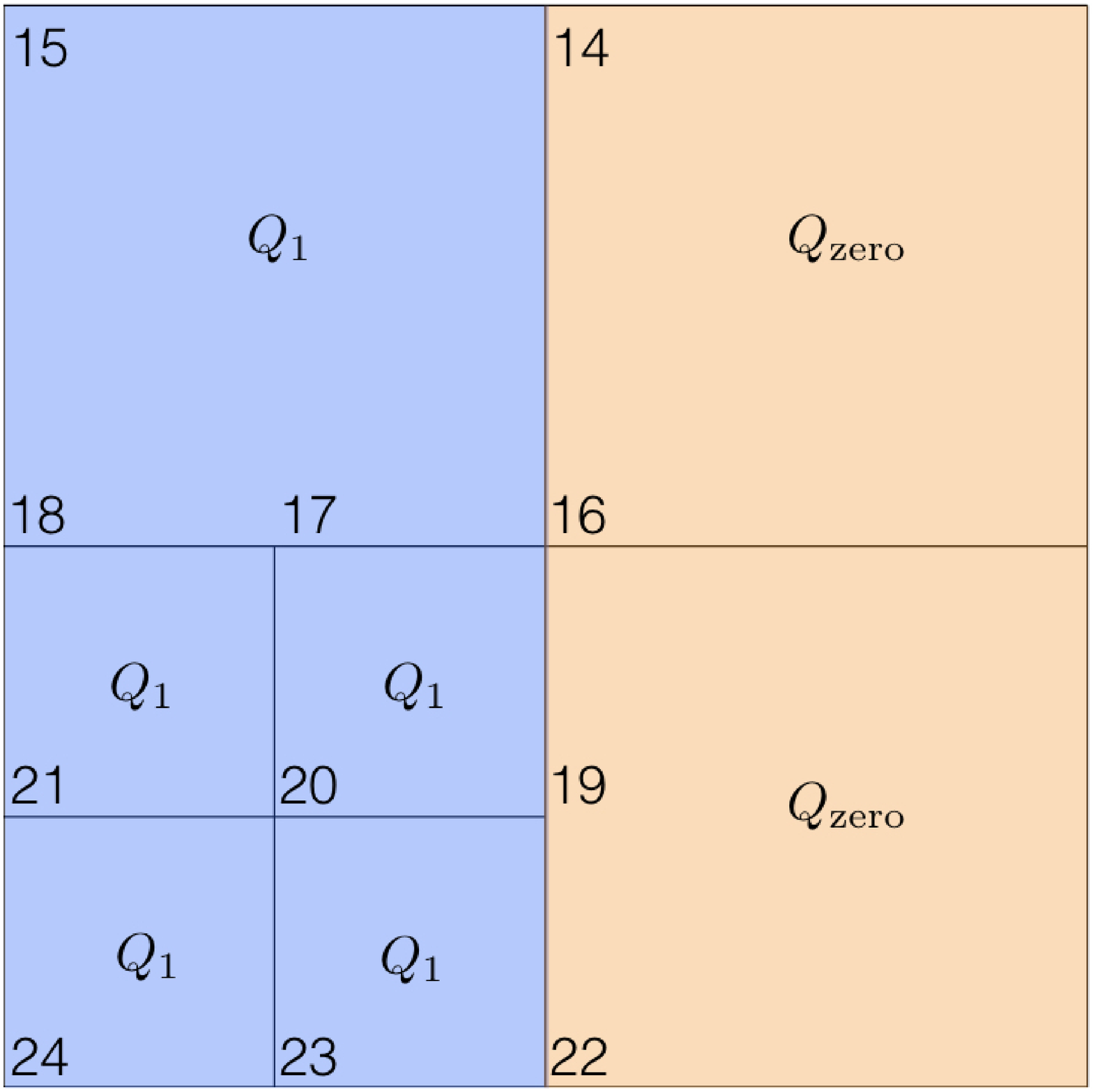}
        \caption{second FE space (enrichment)}
        \label{fig:pou_hanging_b}
    \end{subfigure}  
    \caption{Treatment of hanging nodes for the $h$\,-adaptive PUM. $Q_1$ denotes (bi)linear FE, whereas $Q_{\rm{zero}}$ denotes elements on which functions in the FE space associated with the enrichment function $f_k(\gz x)$ are zero and thus no DoFs need to be introduced.
    }
    \label{fig:pou_hanging}
\end{figure}

An enriched finite element class has been implemented for the general purpose object-oriented \verb!C++! finite element library \texttt{deal.II} \cite{dealII84}. The implementation is based on the \texttt{FESystem} class, which is used to build finite elements for vector valued problems from a list of base (scalar) elements. 
What differs from that class is that the developed FE implementation is scalar, but built from a collection of base elements and  enrichment functions 
\footnote{If we can find a FE space $\widetilde{N}_l$ which contains $N_i$ and $N_{jk}$, then the vector space of (\ref{eq:POUdealii}) is contained in one, built using (\ref{eq:POU}) with $\widetilde{N}_l$. 
    In practice one could use linear shape functions for enriched DoFs and possibly higher order shape functions for non-enriched DoFs.}
\begin{align}
u(\gz x) = \sum_{i \in I} N_i(\gz x) u_i +\sum_{k \in S} f_k(\gz x) \left[\sum_{j \in I^{\rm{pum}}_k} N_{jk}(\gz x) \widetilde{u}_{jk} \right],
\label{eq:POUdealii}
\end{align}
where $I$ is the set of all DoFs with standard shape functions (see Figure \ref{fig:pou_hanging_a}), $I^{\rm{pum}}_k$ is the set of all DoFs corresponding to shape functions enriched with $f_k(\gz x)$ (see Figure  \ref{fig:pou_hanging_b}) and $S$ is the set of enrichment functions. 

As distribution of DoFs in \texttt{deal.II} is element based, we always enrich all DoFs on the element. To restore $C^0$ continuity between enriched and non-enriched elements, additional algebraic constraints are added to force DoFs $\widetilde{u}_{jk}$ associated with $N_{jk} f_k$ on the face between the enriched and non-enriched elements to be zero. This is equivalent to enriching only those shape functions whose support is contained within the enriched elements.

The $h$\,-refinement in \texttt{deal.II} is implemented using hanging nodes. 
In this case, extra algebraic constraints have to be added to make the resulting field conforming. 
We build these constraints separately for the non-enriched FE shape functions and enriched shape functions; 
that is, the following spaces are separately made conforming: $\{ N_i(\gz x)\}$, $\{N_{j0}(\gz x)\}$, $\{N_{j1}(\gz x)\}$, etc. 
To illustrate this idea consider two separate FE spaces shown in Figure \ref{fig:pou_hanging}. 
We assume that functions in the first space are non-zero everywhere in the domain, whereas functions in the second space are non-zero only in the left part, marked by the blue shading. 
Therefore we do not have to introduce any DoFs in the right part, the underlying elements are denoted by $Q_{\rm{zero}}$. 
The standard procedure implemented in \texttt{deal.II} \cite{Bangerth:2008uw} will enforce continuity of the vector field by introducing algebraic constraints for DoFs associated with hanging nodes\footnote{For linear FEs, the value at the hanging node is the average of the values at adjacent vertices, for example $u_5=\nicefrac{1}{2}[u_8+u_2]$.} ($3,5,17,19$), plus constraints for DoFs $14,16,22$ to make functions in the second FE space zero at the interface between $Q_1$ and $Q_{\rm{zero}}$. 
We can observe now that if we take the constrained scalar field from the first FE space and add a scalar field from the second FE space multiplied by the enrichment functions $f(\gz x)$ (continuous in space), the resulting scalar FE field will also be continuous. 
Thus we arrive at a conforming $h$\,-adaptive PUM space where only some elements are enriched. 
With reference to Figure \ref{fig:pou_hanging}, the resulting PUM field will have enrichment associated with DoFs $23,24,20,21,17,18,15$ whereas DoFs $22,19,16,14,17$ will be constrained.

\begin{figure}[!ht]
    \begin{subfigure}[b]{0.20\textwidth}
        \centering
        \includegraphics[width=0.98\textwidth]{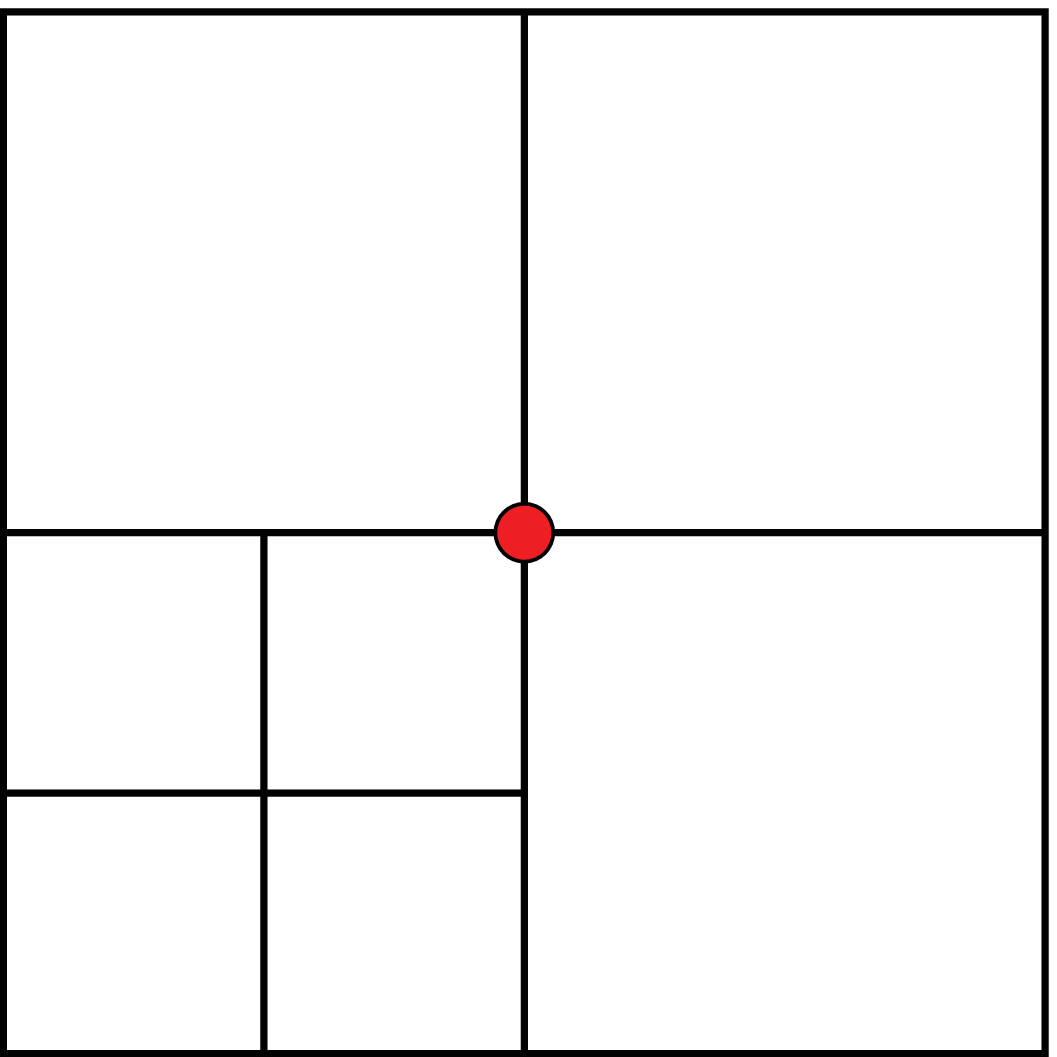}
        \caption{mesh}
    \end{subfigure}  
    ~
    \begin{subfigure}[b]{0.35\textwidth}
        \centering
        \includegraphics[width=0.98\textwidth]{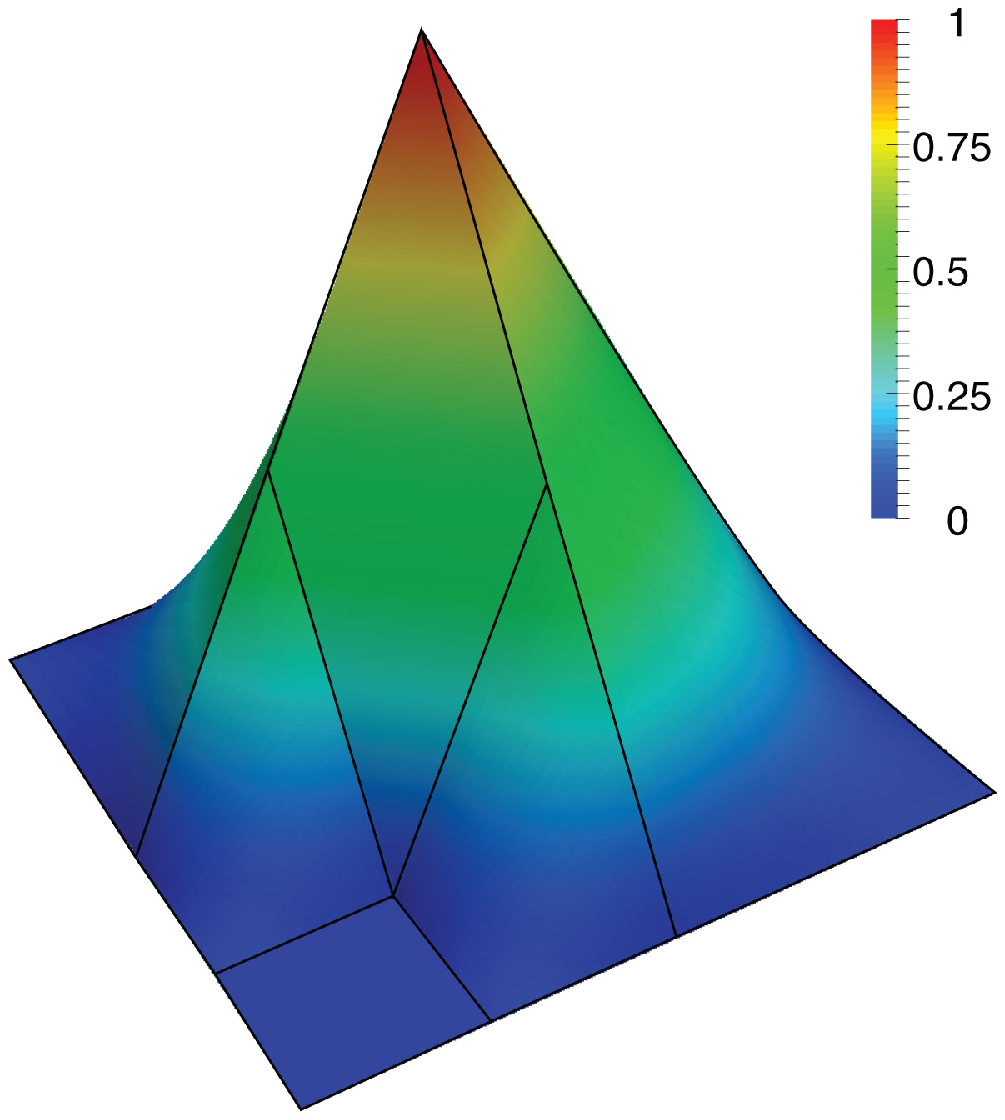}
        \caption{bilinear}
    \end{subfigure}  
    ~
    \begin{subfigure}[b]{0.35\textwidth}
        \centering
        \includegraphics[width=0.98\textwidth]{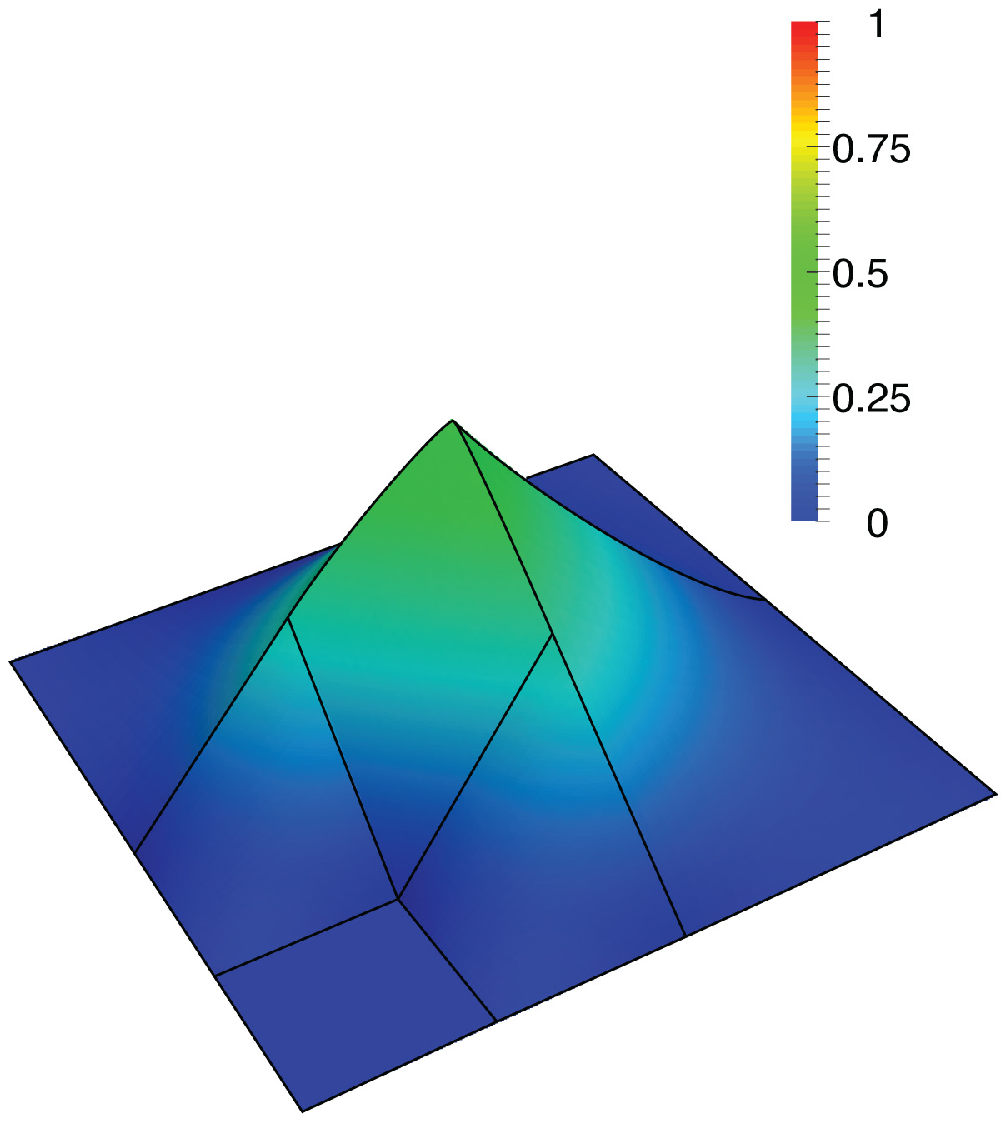}
        \caption{bilinear enriched with $\exp(-\norm{\gz x})$}
    \end{subfigure}
    \caption{$h$\,-adaptive mesh refinement and shape functions associated with the central node on the domain $[0,1]^2$ for the standard and enriched element.
    }
    \label{fig:shape_functions}
\end{figure}

In this procedure the algebraic constraints do not depend on the enrichment functions and are equivalent to those one would have for the vector-value bases build upon the same list of scalar FEs.
Therefore, no extension of the existing functionality to build algebraic constraints was necessary.
This allows us to reuse the code written for the \texttt{FESystem} class.
Figure \ref{fig:shape_functions} depicts an example of enriched and non-enriched shape functions for the case of $h$\,-adaptive refinement with hanging nodes in two dimensions.

\subsection{The choice of enrichment}
Most of the time the \textit{a-priori} knowledge of the solution is limited. In DFT calculation of molecules often core electrons (as opposed to valence electrons) behave in a similar way to single atom solutions. Thus the corresponding solution of single atom problems is used as enrichment functions.
To mimic this in the here considered test eigenproblems, we will only use the lowest eigenvector as an enrichment.
Therefore, for the eigenvalue problems we will employ exponential enrichment.
To lower computational costs we enrich only a subset of elements, chosen based on the input mesh according to vicinity of the element's center to the origin. 

There is another, more important reason why one should limit the enrichment radius. 
There exist combinations of local approximation spaces and partitions of unity that lead to 
linear dependent local basis functions that, consequently, do not form a 
basis of the PUM space \cite{Babuska1997}. The authors in \cite{Babuska1997} give an example of 
piecewise linear hat functions, which form the partition-of-unity, enriched with polynomial local approximation spaces.
In principle these shape functions can still be used but 
the resulting matrices become positive semi-definite 
(as opposed to positive definite).

\begin{figure}[!ht]
    \begin{subfigure}[b]{0.49\textwidth}
        \centering
        \includegraphics[width=0.98\textwidth]{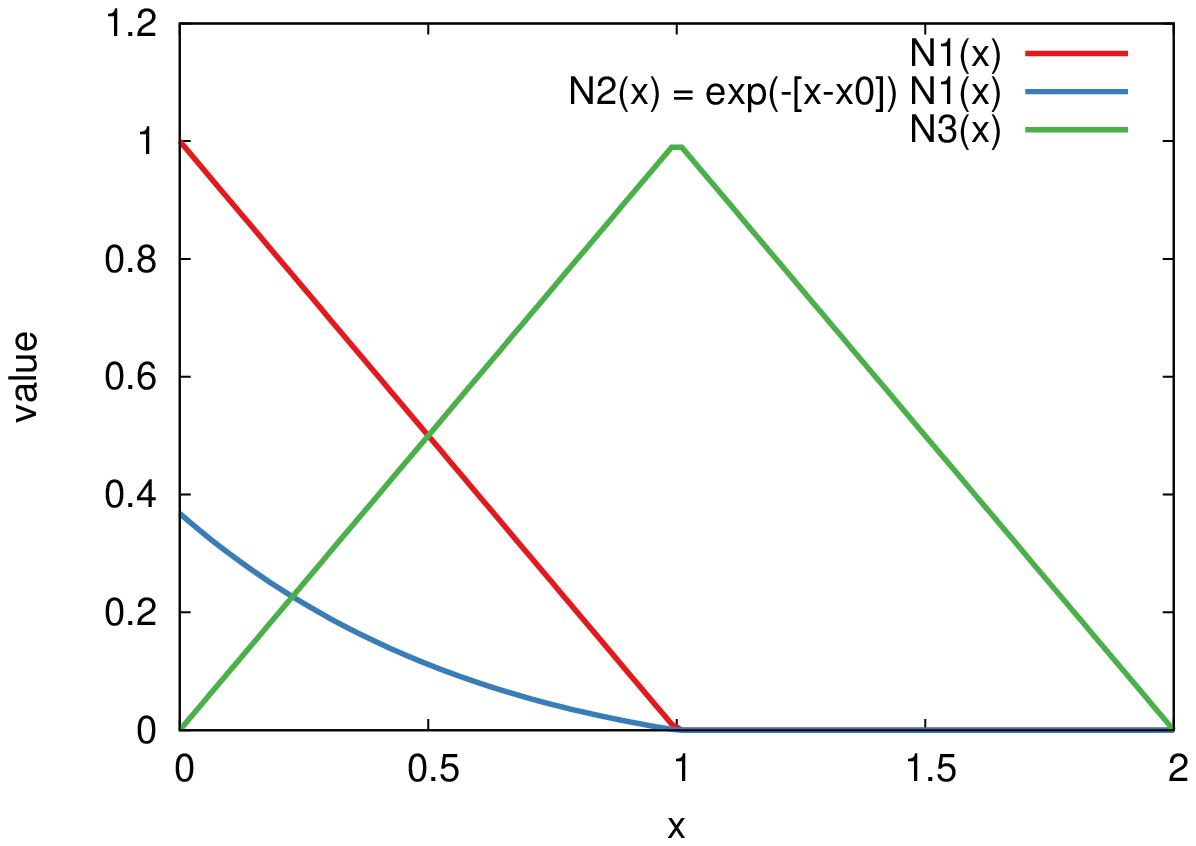}
        \caption{PUM space}
        \label{fig:pou_1d}
    \end{subfigure}  
    ~
    \begin{subfigure}[b]{0.49\textwidth}
        \centering
        \psfrag{x}[c][c]{$x_0$}
        \psfrag{det}[c][c]{$\rm{determinant}$}
        \includegraphics[width=0.98\textwidth]{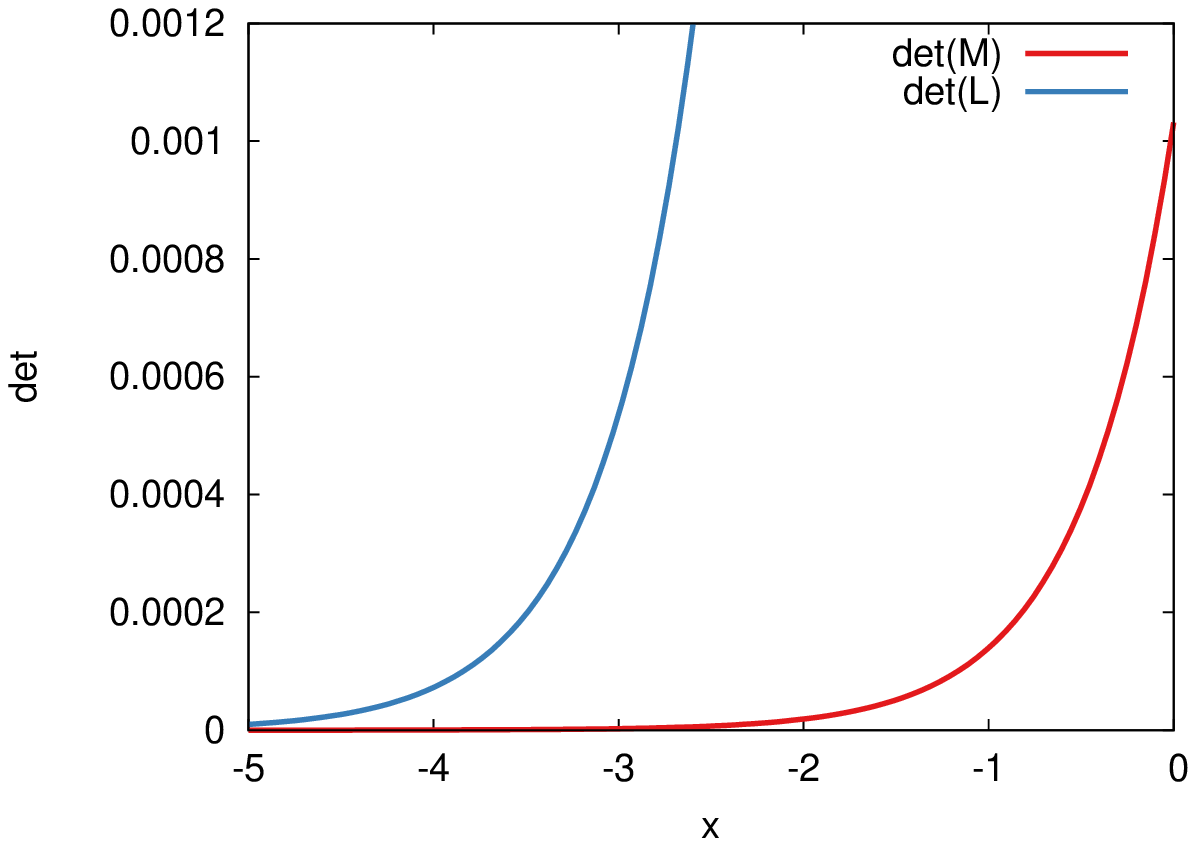}
        \caption{Determinant}
    \end{subfigure}  
    \caption{Enrichment with exponential functions. $x_0$ is the position of the singularity.
    }
    \label{fig:exp_enrichment}
\end{figure}

As a further example, consider a one dimensional mesh with two linear FEs where the first one is enriched with an exponential function, as is shown in Figure \ref{fig:pou_1d}. Determinants of the mass and Laplace matrices quickly tend to zero as the singularity point $x_0$ moves away from the enriched element. Similar behavior can be expected in three dimensions. 

From the practical perspective we notice that when the enrichment radius is too big, the variational convergence of the eigenvalues is lost; that is, the eigenvalues do not necessarily converge from the above to the exact values. 
To avoid such behavior the radius of enrichment has to be limited. 
The exact radius is contingent upon the decay of the enrichment function and the initial mesh. 
Note that in \cite{Sukumar:2009vw} enrichment for the harmonic oscillator problem is also localized to a predefined maximum distance. 
The authors, however, do not discuss the rationale for their choice of cut-off radius.

\subsection{Numerical integration}
One of the particular features of PUM that needs careful treatment is numerical integration.
Integrands in the weak form become less smooth and attaining a higher accuracy of the integration is therefore a more difficult task.
There are several approaches to address this. 
One is adaptive integration schemes (for example \cite{Mousavi2012}), when the element over which the integration is performed is subdivided into child elements iteratively until the convergence of the integral is attained. 
Implementation of this procedure in the \texttt{deal.II} library is, unfortunately, not straightforward.
An alternative is to perform coordinate transformation, such as the cubic transformation proposed in \cite{Telles1987}. However the generalization of this approach to 3D appears to require rectangular hexahedron elements, which would be a major constraint in generating input meshes.
As a result, similar to \cite{Patzak2003} we have opted to utilize higher order Gaussian quadrature rules. 
For the numerical results presented below this approach was found to produce sufficiently accurate results while not becoming a bottleneck in calculations.

\subsection{Error estimator}
\textit{A posteriori} error estimation analysis for FE approximations of (second-order) eigenvalue problems has been a topic of intensive study within the last several decades, both from theoretical and implementational standpoints. We refer the interested reader to \cite{Verfuerth1996,Larson2000,Heuveline2001,Duran2003,Mao2006,Dai2008,Garau2009}, where both residual- and averaging-based error estimators are presented.

Let $\{\psi^h,\lambda^h\}$ denote the set of eigenpairs computed on a finite element mesh $\mathcal{P}^h$.
In general, a discretization error in approximated eigenfunctions, $\psi-\psi^h$, measured in a suitable norm (e.g. $L^2$ and energy norm), as well as in approximated eigenvalues, $|\lambda-\lambda^h|$, can be estimated from above. 
That is, 
    \begin{equation}
    \left\|\psi-\psi^h\right\|\leq C_1\eta,
    \label{bound_eigenvectors}
    \end{equation}
    and
    \begin{equation}
    |\lambda-\lambda^h|\leq C_2\eta^2,
    \label{bound_eigenvalues}
    \end{equation}
    where $C_1,C_2$ are stability constants that are independent of the mesh size and $\eta$ is the {\em explicitly computable} error upper-bound, see e.g. \cite{Larson2000,Dai2008} for details. These equations are typically termed {\em (global) error estimators}. The bound $\eta$ reads as 
    \begin{equation*}
    \eta:=\left[\sum_{K\in\mathcal{P}^h} \eta_K^2 \right]^\frac{1}{2},
    \end{equation*}
    where summation is performed over all elements in $\mathcal{P}^h$ and $\eta_K$ is the {\em (local) error indicator},
    a quantity showing a discretization error of $\{\psi^h,\lambda^h\}$ element-wise, that is, on every fixed $K$. With multiple solutions available (in this case, eigenpairs $\{\psi^h_\alpha,\lambda^h_\alpha\}$), $\eta_K$ will be a sum of discretization errors of the corresponding eigenpairs on a given element $K$, that is
\begin{equation*}
\eta_K:=\left[\sum_{\alpha} \eta_{K,\alpha}^2 \right]^\frac{1}{2}.
\end{equation*}

For a standard (non-enriched) $\mathbb{Q}_1$-based finite element solution of (\ref{eq:Schroedinger}), a local indicator $\eta_{K,\alpha}$ of so-called {\em residual} type reads as follows (see \cite{Larson2000,Heuveline2001,Dai2008,Garau2009} for details):
\begin{align}
\begin{split}
\eta_{K,\alpha}^2
:= 
\,h_K^2 \int_K \left[ \,\Bigg( -\frac{1}{2}  \nabla^2 + V({\bf x})\Bigg) \psi^h_\alpha- \lambda^h_\alpha\,\psi^h_\alpha  \right]^2 \rm d \gz{x}  
+
\, h_K \sum_{e\subset\partial K}  \int_e \dbracket{-\frac{1}{2}\gz \nabla \psi^h_\alpha \cdot \gz n}_e^2 \rm d \gz a,
\end{split}
\label{eq:residual_indicator}
\end{align}
where $[\![-\frac{1}{2}\gz \nabla \psi^h_\alpha \cdot \gz n ]\!]_e:=\left[-\frac{1}{2} \gz \nabla \psi^h_\alpha\,|_{K}+\frac{1}{2}\nabla \psi^h_\alpha\,|_{K'} \right]\cdot \gz n_e$ represents the jump of the gradient across interface $e$ between two adjacent elements $K$ and $K'$, $\gz n_e$ is the outward unit normal vector to $e$ and $h_K:=\mathrm{diam}(K)$.

One of the findings of our work is the proof that indicator (\ref{eq:residual_indicator}) can also be used in the PUM with the exponential enrichment function $f(\gz x) = \exp{(-\mu \norm{\gz x}^p)}$. In the appendix, we derive and prove the related local interpolation error estimates required for the derivation of the error estimator in this case.

\section{hp-adaptive solution}
\label{sec:hp}

There have been numerous works devoted to $hp$\,-adaptive refinement \cite{Houston2003, Melenk2001, Heuveline2003, Houston2005, Hartmann2010, Mavriplis1994, Eibner2007} including a comparison of different methods \cite{Mitchell2014}.
The main difficulty that \textit{a posteriori} $hp$\,-adaptive methods aim to address is the following: Once an error is estimated and a certain subset of elements is marked for refinement, one has to choose between $h$\,- or $p$\,-refinement for each element.
It is a general knowledge that it is better to increase polynomial degree ($p$\,-refinement) of those elements where the solution is smooth, 
whereas it is better to refine the element ($h$\,-refinement) near the singularities of the solution.

In this work we adopt a strategy based on the estimate of the analyticity of the solution\footnote{that is the measure of how well it is representable by power series} on the reference element via expansion into a Legendre basis \cite{Houston2005, Hartmann2010, Mavriplis1994, Eibner2007}.
In particular, we perform a least squares fit of Legendre coefficients $a_i$ for each element 
\begin{align}
\norm{a_i} \sim C \exp(-\sigma i) .
\end{align}
The minimum decay coefficient $\sigma$ in each direction is used to estimate analyticity as $\exp(-\sigma)$.
This corresponds to an estimation of smoothness in the direction where the solution is roughest. 
As there is no anisotropic elements in \texttt{deal.II} that can be used with $hp$\,-refinement, distinguishing between different directions is not needed.
When this value is below $\exp(-1)$, the solution is considered to be smooth and thus $p$\,-refinement is performed, otherwise $h$\,-refinement is executed. 
For linear FEs $p$\,-refinement is always performed. 
Finally, in order to avoid numerical issues with the evaluation of $\log \norm{a_j}$, 
for the least squares fit
we only consider coefficients that are 
two orders of magnitude greater than the minimal representable positive floating value.

In order to extend this $hp$\,-refinement strategy to the eigenvalue problem, that is when there are multiple vectors represented using the same FE basis, we propose the following approach: 
For each element we find an eigenvector which contributes the most to the total element's error. 
The smoothness of this vector is the basis on which we decide to perform $h$\,-refinement or $p$\,-refinement. 
The rationale behind this approach is that we aim at minimizing the error the most during a single refinement step while being conservative and avoiding performing both $h$\,- and $p$\,-refinement on the same element. 
In our opinion the proposed strategy is a better choice than, for example, choosing minimum smoothness among all vectors for a given element. 
That may be considered to be a more robust approach but could also lead to a slower global convergence.

Finally, for the error indicator we adopt the following expression \cite{Giani2012}
\begin{align}
\eta^2_{K,\alpha} := 
\frac{h_{K}^2}{p_K^2} \int_{K} \left[ \,\Bigg( -\frac{1}{2}  \nabla^2 + V({\bf x})\Bigg) \psi^h_\alpha- \lambda^h_\alpha\,\psi^h_\alpha  \right]^2 \d \gz x + 
\sum_{e \subset \partial K} \frac{h_{e}}{2 p_e} \int_{e} \dbracket{-\frac{1}{2}\gz \nabla \psi_\alpha^h \cdot \gz n }_e^2 \d \gz a \,,
\label{eq:residual_indicator_hp}
\end{align}
where 
$h_e$ is the face's diameter, 
$p_K$ is the element's polynomial degree and $p_e$ is the maximum polynomial degree over two elements $K$ and $K'$ adjacent to the face $e$.

\section{Results and Discussion}
\label{sec:results}

If not explicitly stated otherwise, the results below are obtained for
the following configuration:
\begin{enumerate*}[label=(\roman*)]
    \item the initial polynomial degree for non-enriched DoFs is one for $hp$\,-adaptive FEM; 
    \item linear shape functions are used for the PUM;
    \item a Gaussian quadrature rule with $20^3$ points is used for enriched elements in the eigenvalue problem;
    \item a Gaussian quadrature rule with $[7+p]^3$ points is used for standard elements in the eigenvalue problem, where $p$ is the polynomial degree of the basis;
    \item the D{\"o}rfler marking strategy with $\theta = 0.6$ is used to mark elements for refinement;
    \item integration of the jump of fields over faces in error estimators
    is performed with $[1+p]^2$ Gaussian quadrature points, 
    where $p$ is the polynomial degree of the basis; 
    \item we assume a Q1 mapping for elements;
    \item Gauss-Legendre-Lobatto supports points are used for the $hp$\,-adaptive FEM basis to improve the condition number;
    \item a standard residual-type error estimator similar to (\ref{eq:residual_indicator}) is used for the Poisson problem in $h$\,-adaptive FEM and PUM calculations
    $\eta^2_{K} :=  h_{K}^2 \int_{K} \left[ \nabla^2 \phi + 4 \pi \rho \right]^2 \d \gz x + h_K \sum_{e \subset \partial K} \int_{e} [\![\gz \nabla \phi \cdot \gz n ]\!]^2_e \d \gz a $;
    \item linear shape functions are used for the FEM and PUM when applied to the Poisson problem;
    \item Parallel vectors, matrices and solvers for linear algebra
    problems in the Portable, Extensible Toolkit for Scientific
    Computation (PETSc) \cite{petsc-user-ref} and parallel solvers
    for eigenvalue problems in the Scalable Library for Eigenvalue Problem Computations (SLEPc) \cite{Hernandez2005} are used for the eigenvalue problem;
    \item Trillinos \cite{Heroux2005} vectors,
    matrices, solvers and preconditioners are used to solve the extended
    Poisson problem.
\end{enumerate*}

In case of $hp$\,-adaptive refinement the highest polynomial degree is limited to $4$. The rationale for that choice is as follows: In order to preserve variational convergence when solving a coupled eigenvalue and Poisson problem in DFT, the polynomial degree of the Poisson FE basis should be twice of that used for the eigenvalue problem \cite{Davydov2014}. Thus quartic FEs in the eigenvalue problem would require polynomials up to $8$-th order in the Poisson problem. From our experience (not reported here) this is already challenging both from the number of DoFs as well as the condition number of the Laplace matrix in $hp$\,-adaptive refinement.

\subsection{Eigenvalue problem}
The initial mesh used to solve the Schr{\"o}dinger equation is obtained from 3 global mesh refinements of the single element in $\Omega=[-20;20]^3$ for the Coulomb potential and $\Omega=[-10;10]^3$ for the harmonic potential. For the PUM only 8 elements adjacent to the singularity that is located at the origin are marked for enrichment.

\begin{figure}[!ht]
    \centering
    \begin{subfigure}[b]{0.49\textwidth}
        \psfrag{DoF}[t][t][0.8]{$\rm{DoFs}$}
        \psfrag{error}[B][B]{$\lambda_0^h-\lambda_0$}
        \includegraphics[width=0.98\textwidth]{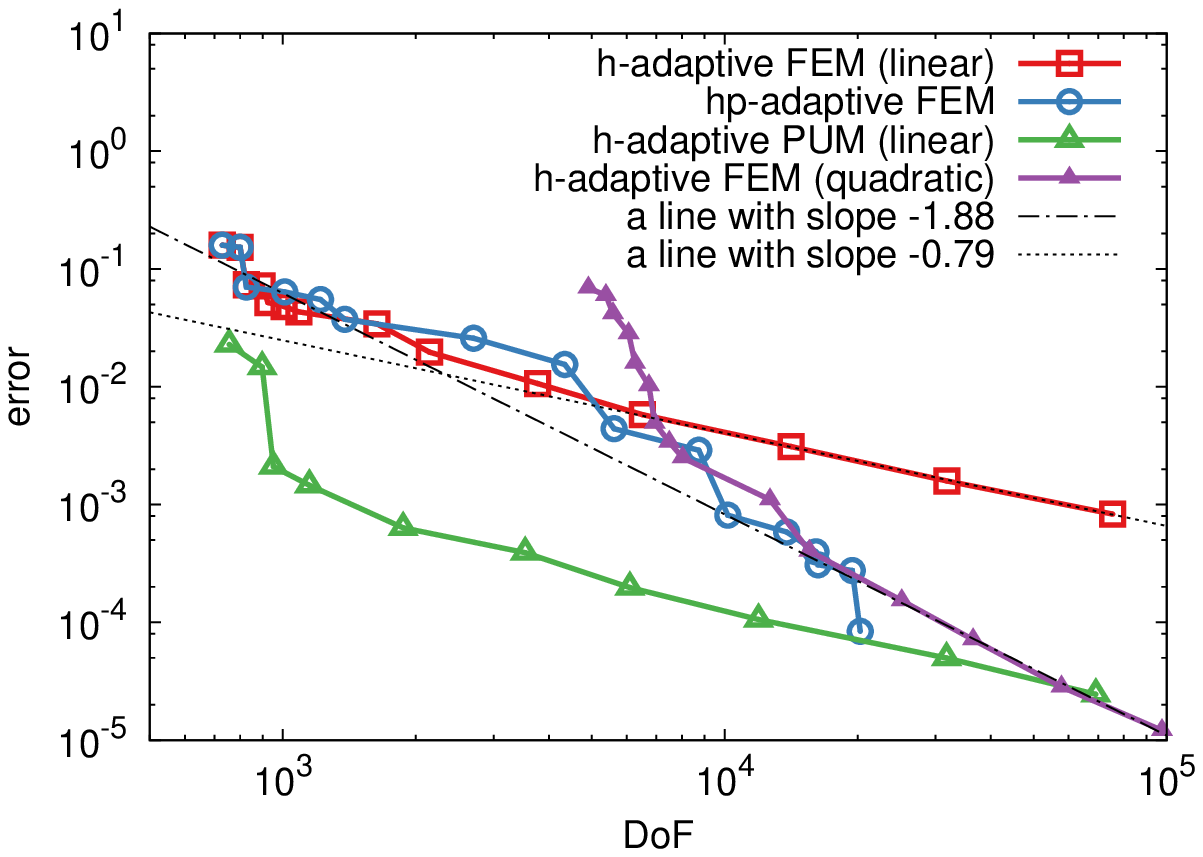}
        \caption{
            Coulomb potential
        }%
        \label{fig:GHEP1_C}
    \end{subfigure}
    \begin{subfigure}[b]{0.49\textwidth}
        \psfrag{DoF}[t][t][0.8]{$\rm{DoFs}$}        \psfrag{error}[B][B]{$\lambda_0^h-\lambda_0$}
        \includegraphics[width=0.98\textwidth]{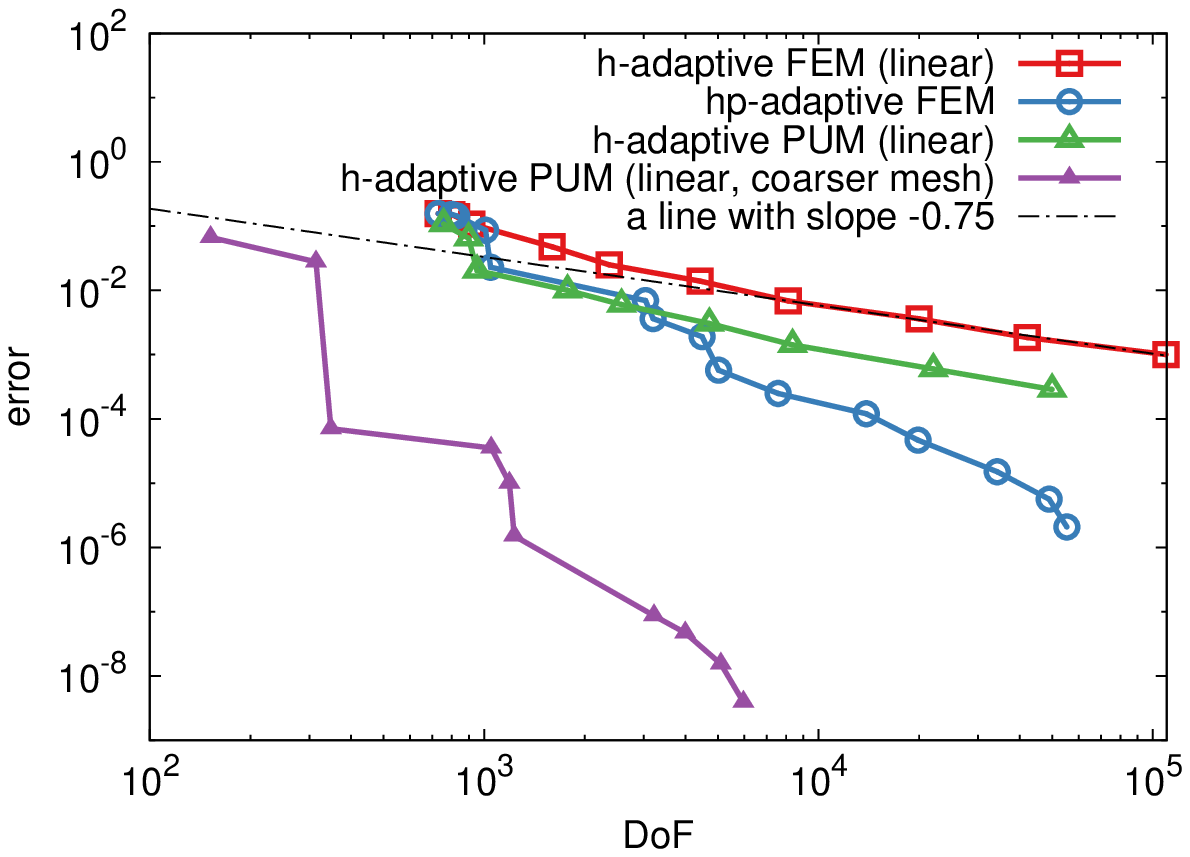}
        \caption{
            Harmonic potential
        }%
        \label{fig:GHEP1_H}
    \end{subfigure}
    \caption{Solving an eigenproblem for a single eigenpair. }
    \label{fig:GHEP1}
\end{figure}

\begin{figure}[!ht]
    \centering
    \begin{subfigure}[b]{0.49\textwidth}
        \includegraphics[width=0.98\textwidth]{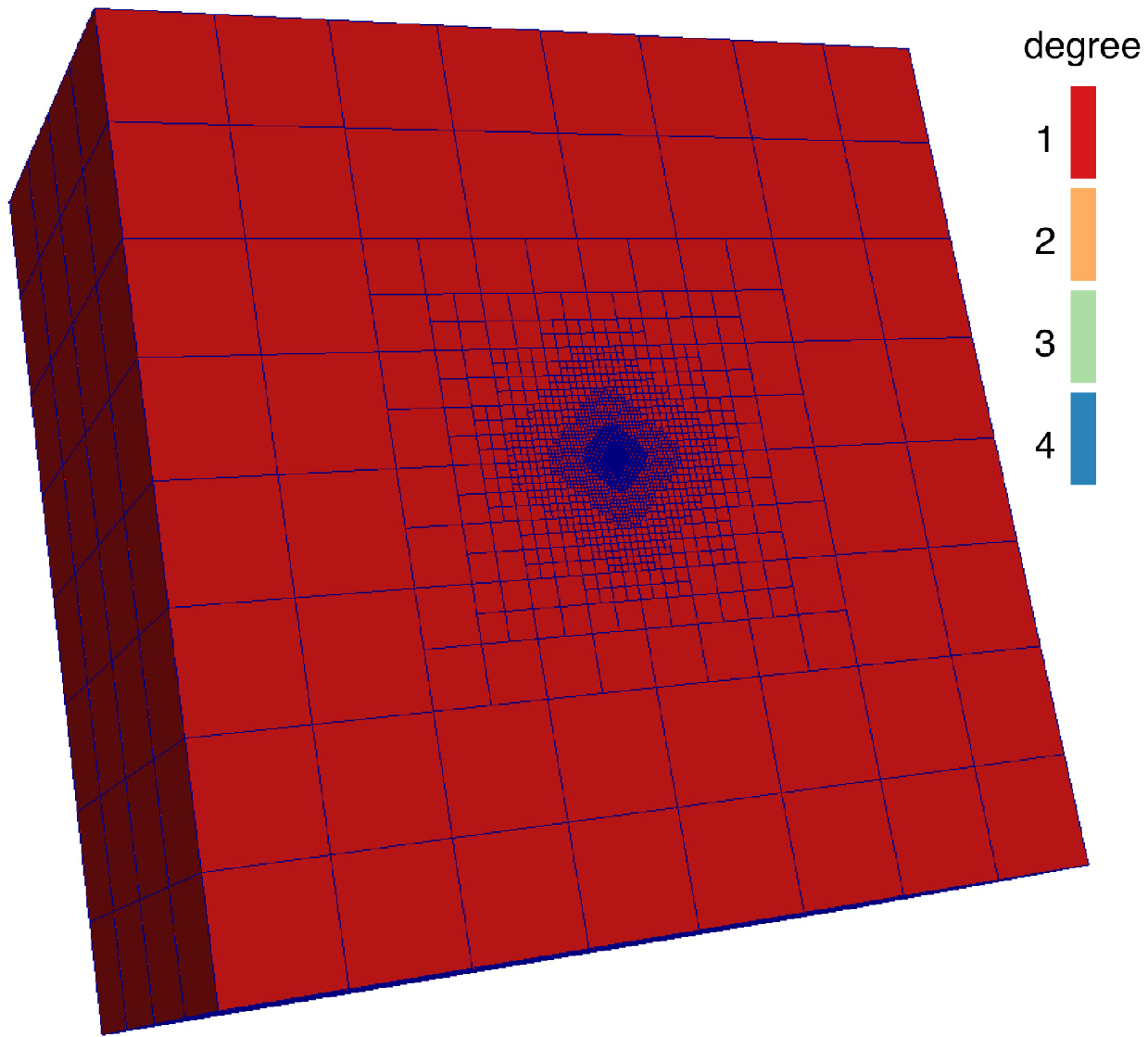}
        \caption{
            $h$\,-adaptive FEM (linear)
        }%
        \label{fig:GHEP1_Ch}
    \end{subfigure}
    \begin{subfigure}[b]{0.49\textwidth}
        \includegraphics[width=0.98\textwidth]{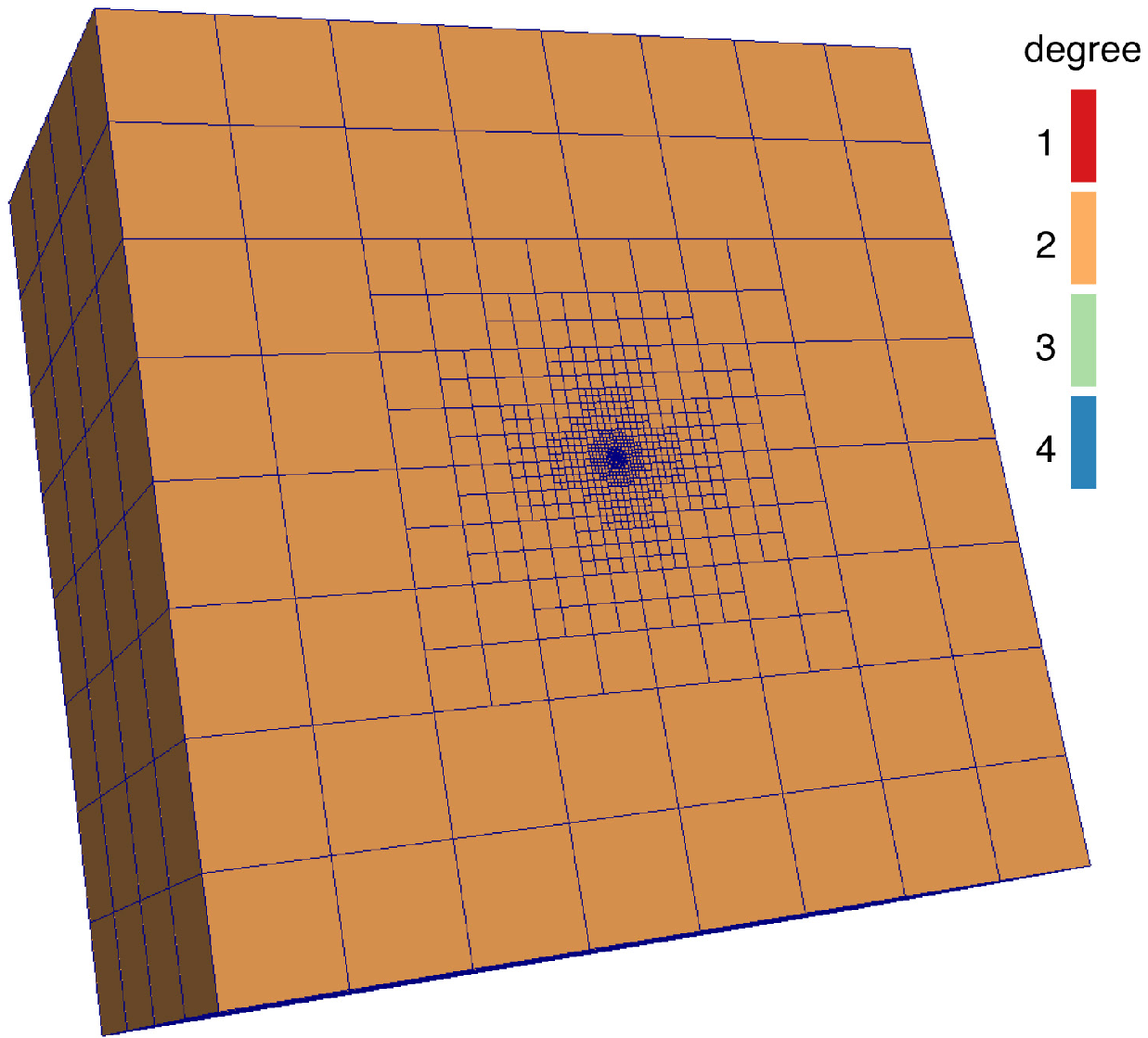}
        \caption{
            $h$\,-adaptive FEM (quadratic)
        }%
        \label{fig:GHEP1_Ch2}
    \end{subfigure}
    \begin{subfigure}[b]{0.49\textwidth}
        \includegraphics[width=0.98\textwidth]{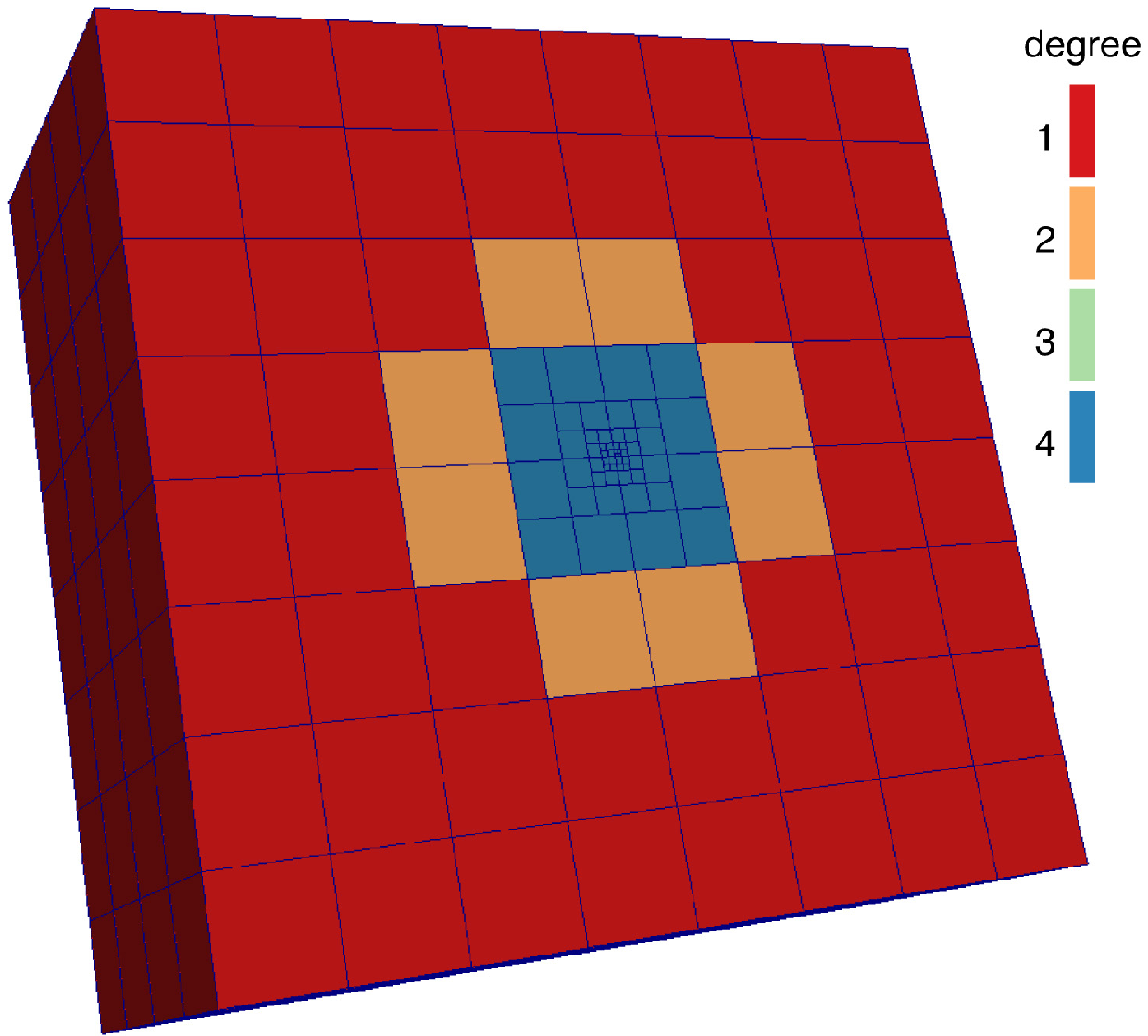}
        \caption{
            $hp$\,-adaptive FEM
        }%
        \label{fig:GHEP1_Chp}
    \end{subfigure}
    \begin{subfigure}[b]{0.49\textwidth}
        \includegraphics[width=0.98\textwidth]{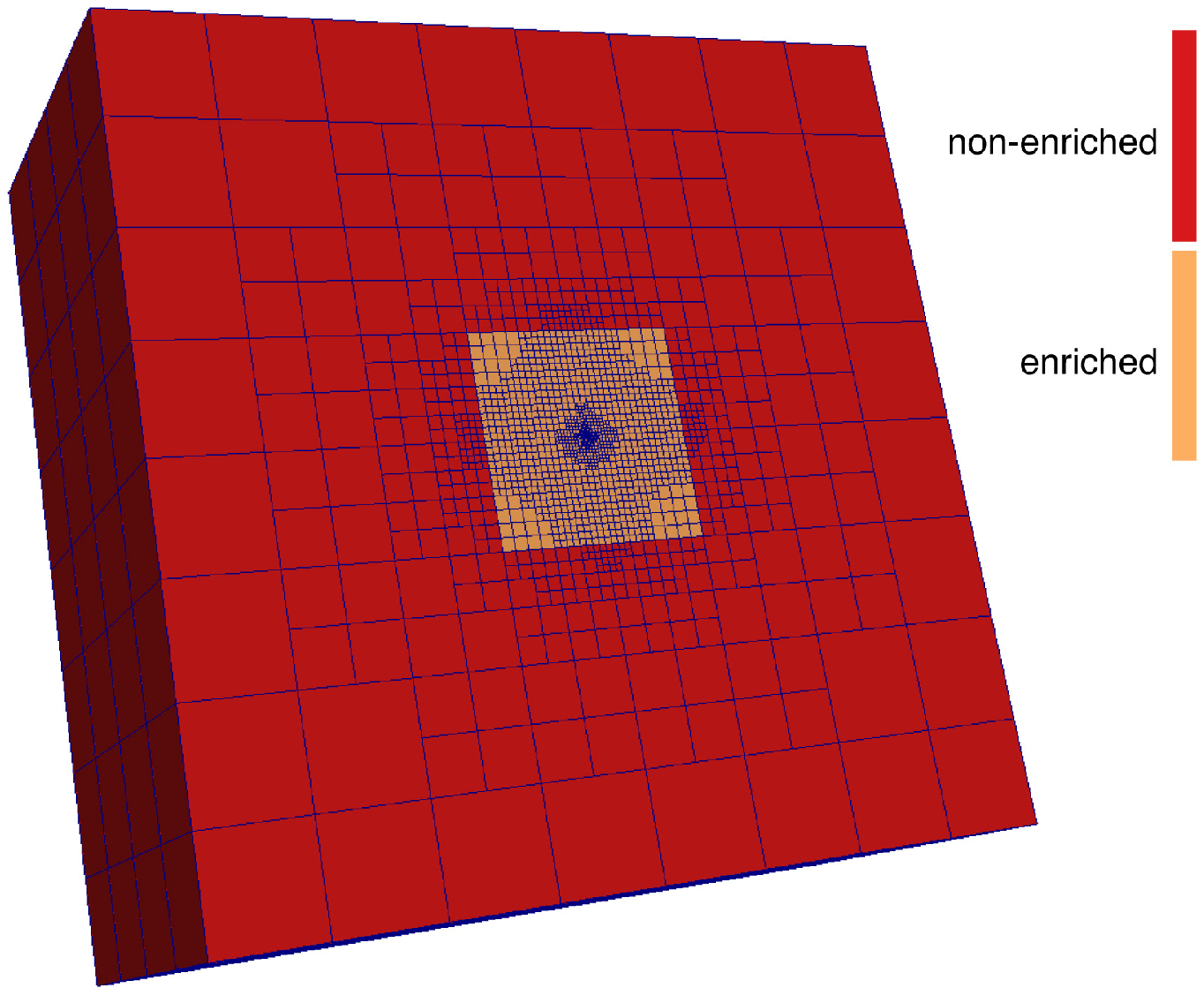}
        \caption{
                $h$\,-adaptive PUM (linear)
        }%
        \label{fig:GHEP1_Cpou}
    \end{subfigure}  
    \caption{Cross-sections of the final meshes for the Coulomb potential when solving for a single eigenpair. 
    }
    \label{fig:GHEP1_meshes}
\end{figure}

First, we examine the convergence in case when a single eigenpair is required in the Schr{\"o}dinger equation with two different potentials.
Figure \ref{fig:GHEP1} compares the $h$\,-adaptive FEM, $hp$\,-adaptive FEM and $h$\,-adaptive PUM, whereas Figure \ref{fig:GHEP1_meshes} shows the cross-sections of meshes for the last refinement step.

For both combination of potentials and enrichment functions, the $h$\,-adaptive PUM is superior to $h$\,-adaptive FEM.
In particular, for the last refinement step the PUM solution is about 2 orders more accurate than the $h$\,-adaptive FEM with the same number of DoFs in case of the Coulomb potential.
For the harmonic potential this value is smaller.
The asymptotic convergence rate of the $h$\,-adaptive PUM with the default enrichment radius is very similar to that of the $h$\,-adaptive FEM for both problems (compare green and red lines in Figure\, \ref{fig:GHEP1}).

The advantage of $h$\,-adaptive PUM also depends on the enrichment radius with respect to the underlying exact solution. 
To examine this effect we employ an initial mesh obtained only by 2 global refinements of a single element and mark the 8 elements adjacent to the origin for enrichment.
With this approach we effectively consider a larger enrichment domain $[-5;5]^3$ instead of $[-2.5;2.5]^3$.
Importantly, the numerically non-zero part of the underlying analytical solution will be almost fully contained in those 8 elements (see Figure \,\ref{fig:schroedinger_solution_H}).
From the numerical results we observe that for the most refined stage the $h$\,-adaptive PUM displays an error which is about 6 orders of magnitude less than the same method with the smaller enrichment domain (compare purple and green lines in Figure \,\ref{fig:GHEP1_H} ).

Interestingly, the $hp$\,-adaptive FEM does not display a big advantage over the $h$\,-adaptive quadratic FEM for the Hydrogen atom and the smoothness estimator considered here (compare blue and purple lines in Figure \,\ref{fig:GHEP1_C}).

\begin{figure}[!ht]
    \begin{subfigure}[b]{0.99\textwidth}
        \centering
        \psfrag{DoF}[t][t][0.8]{$\rm{DoFs}$}
        \psfrag{error}[B][c]{$\lambda_{\alpha}^h-\lambda_{\alpha}$}        
        \includegraphics[width=0.98\textwidth]{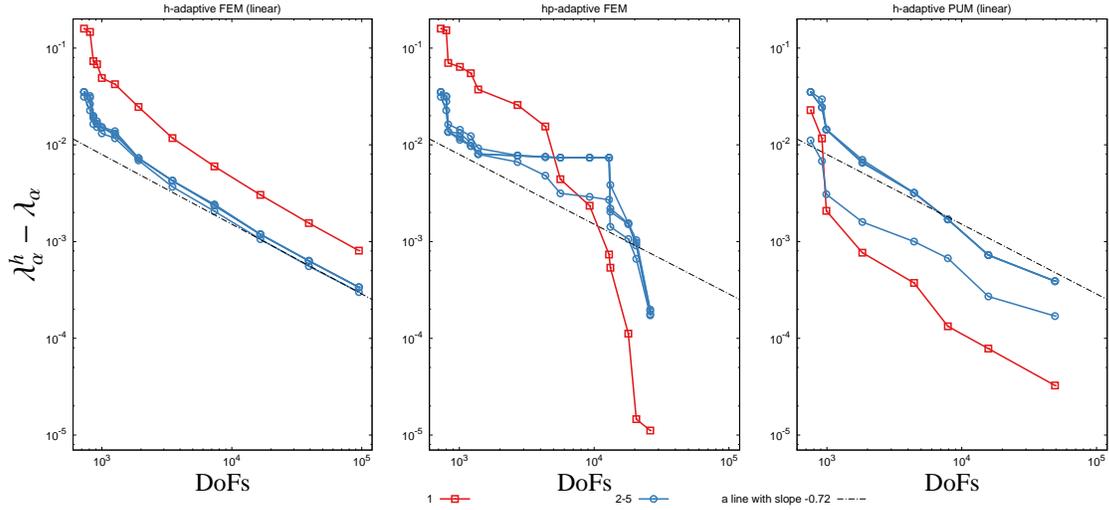}
        \caption{ 
            Coulomb potential (4 out of 5 eigenvalues are degenerate).
        }%
        \label{fig:CHEP_mult_C}
    \end{subfigure}
    ~
    \begin{subfigure}[b]{0.99\textwidth}
        \centering
        \psfrag{DoF}[t][t][0.8]{$\rm{DoFs}$}
        \psfrag{error}[B][c]{$\lambda_{\alpha}^h-\lambda_{\alpha}$}        
        \includegraphics[width=0.98\textwidth]{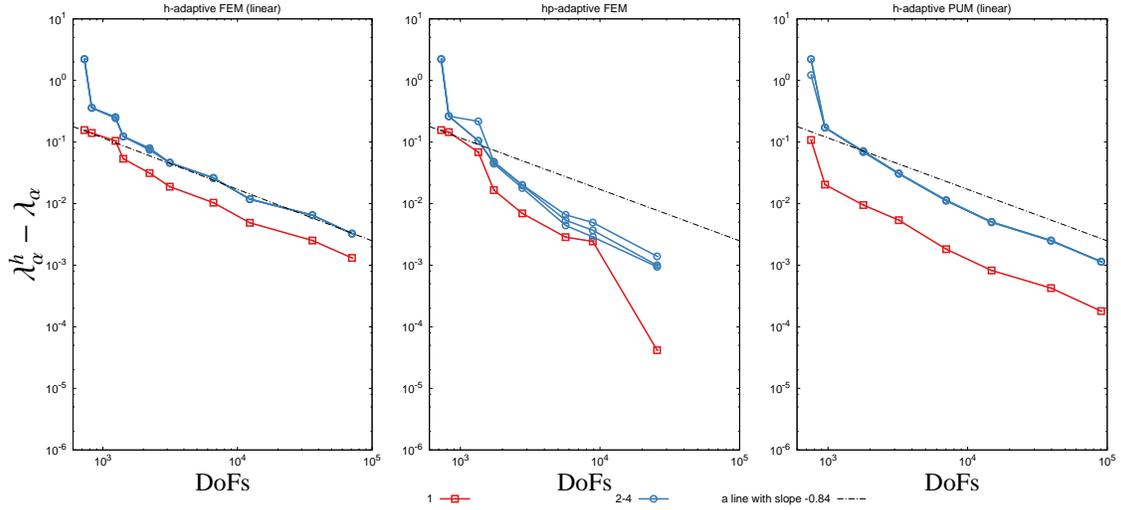}
        \caption{ 
            Harmonic potential (3 out of 4 eigenvalues are degenerate).
        }%
        \label{fig:CHEP_mult_H}
    \end{subfigure}
    \caption{Convergence of eigenvalues from the first two energy levels for the Schr{\"o}dinger equation in the course of adaptive refinement. Red lines denote the lowest eigenvalue, whereas blue lines correspond to degenerate eigenvalues on the next energy level.}
    \label{fig:CHEP_mult}
\end{figure}

Now let us turn our attention to a more realistic scenario where one seeks multiple eigenpairs whereas an \textit{a priori} knowledge is available only for the first eigenvector.
Figure \ref{fig:CHEP_mult} plots convergences of the first 5 / 4 eigenvalues for the Schr{\"o}dinger equation with Coulomb / harmonic potential solved with the different methods.
For both problems the $h$\,adaptive PUM again has remarkable convergence properties, superior to 
$h$\,-adaptive FEM.
It is important to note that even though in the PUM the enrichment function corresponds to the first eigenvector only, others eigenpairs in the case of the harmonic potential tend to converge faster than the standard $h$\,-adaptive FEM case, as can be observed in Figure \ref{fig:CHEP_mult_H}. 
The same applies to the spherical orbital at the second energy level of the Hydrogen atom; see Figure \ref{fig:CHEP_mult_C} where the corresponding eigenvalue in the PUM case displays a faster convergence rate than the others on the same energy level.

For the Hydrogen atom,
in the case of the $hp$\,-adaptive refinement one observes a superior convergence rate of the first eigenvalue, whereas eigenvalues from the next energy level have higher errors at some stages when compared to $h$\,-adaptive linear FEM. 
This indicates that the suggested strategy of deciding between $h$\,- or $p$\,-refinement for multiple degenerate eigenvectors is not ideal. 
A possible issue could be related to smoothness estimation on elements with hanging nodes.
In particular it is observed \cite{dealii-tutorial27} that the smoothness is overestimated when using similar methods, albeit based on Fourier coefficients. This leads to unnecessarily high order polynomial degrees in these areas. 

In DFT calculations, the requested tolerance of eigenvalues is often at the order of $10^{-3}$. 
In this case, it is clear from Figure \ref{fig:CHEP_mult_H} that for the Hydrogen atom the linear PUM achieves this tolerance for all eigenvalues at a number of DoFs comparable to the $hp$\,-adaptive FEM. 
Thus, depending on the required accuracy, the $h$\,-adaptive PUM can be as efficient as the $hp$\,-adaptive FEM.

\subsection{Poisson problem}
In this subsection we turn our attention to the solution of the Poisson problem with the physical interpretation here being the electrostatic potential produced by the charge density. We will consider the solution obtained for two different values of the regularization parameter $\sigma$, namely $1.0$ and $0.1$ (their influence is shown in Figure \ref{fig:density_solution}).

\begin{figure}[!ht]
    \begin{subfigure}[b]{0.48\textwidth}
        \centering
        \psfrag{DoF}[t][t][0.8]{$\rm{DoFs}$}
        \includegraphics[width=0.98\textwidth]{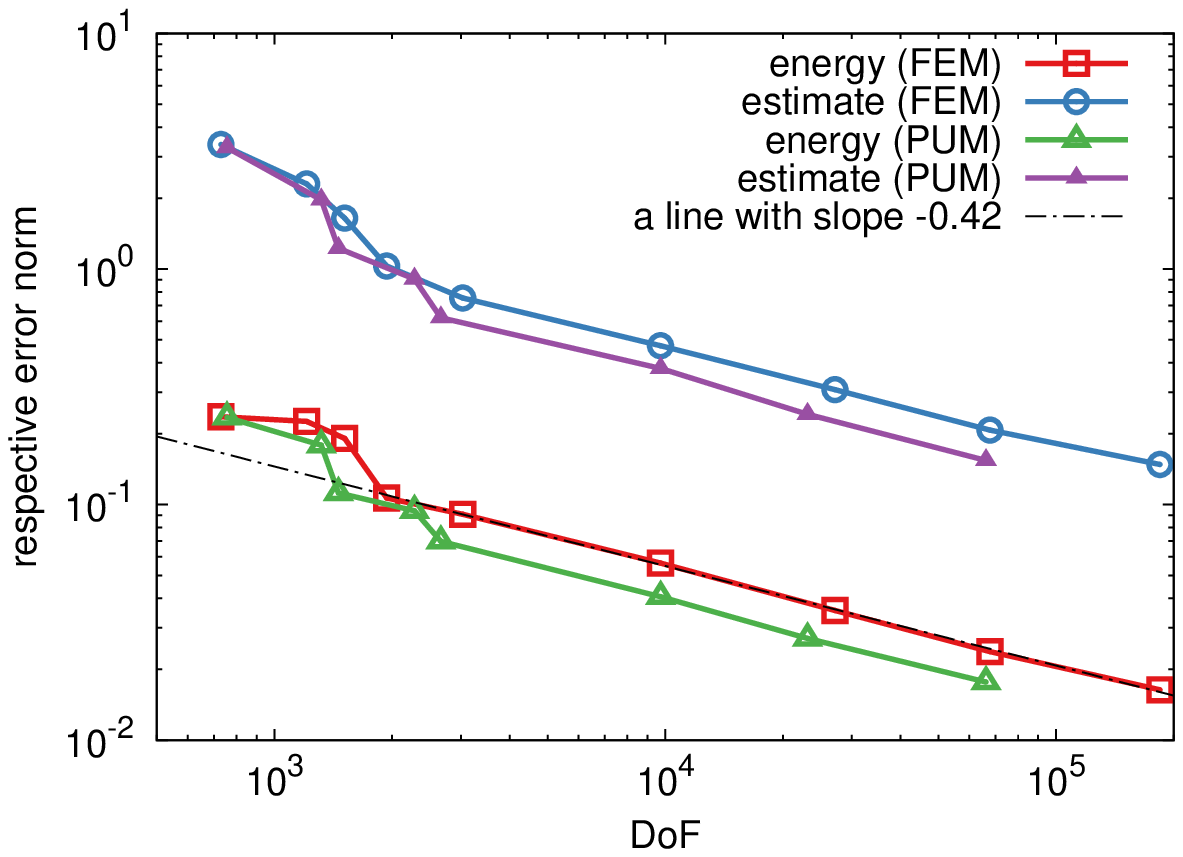}
        \caption{ 
            $h$\,-adaptive.
        }%
        \label{fig:Poiss_HG_h}
    \end{subfigure}
    ~
    \begin{subfigure}[b]{0.48\textwidth}
        \centering
        \psfrag{DoF}[t][t][0.8]{$\rm{DoFs}$}
        \includegraphics[width=0.98\textwidth]{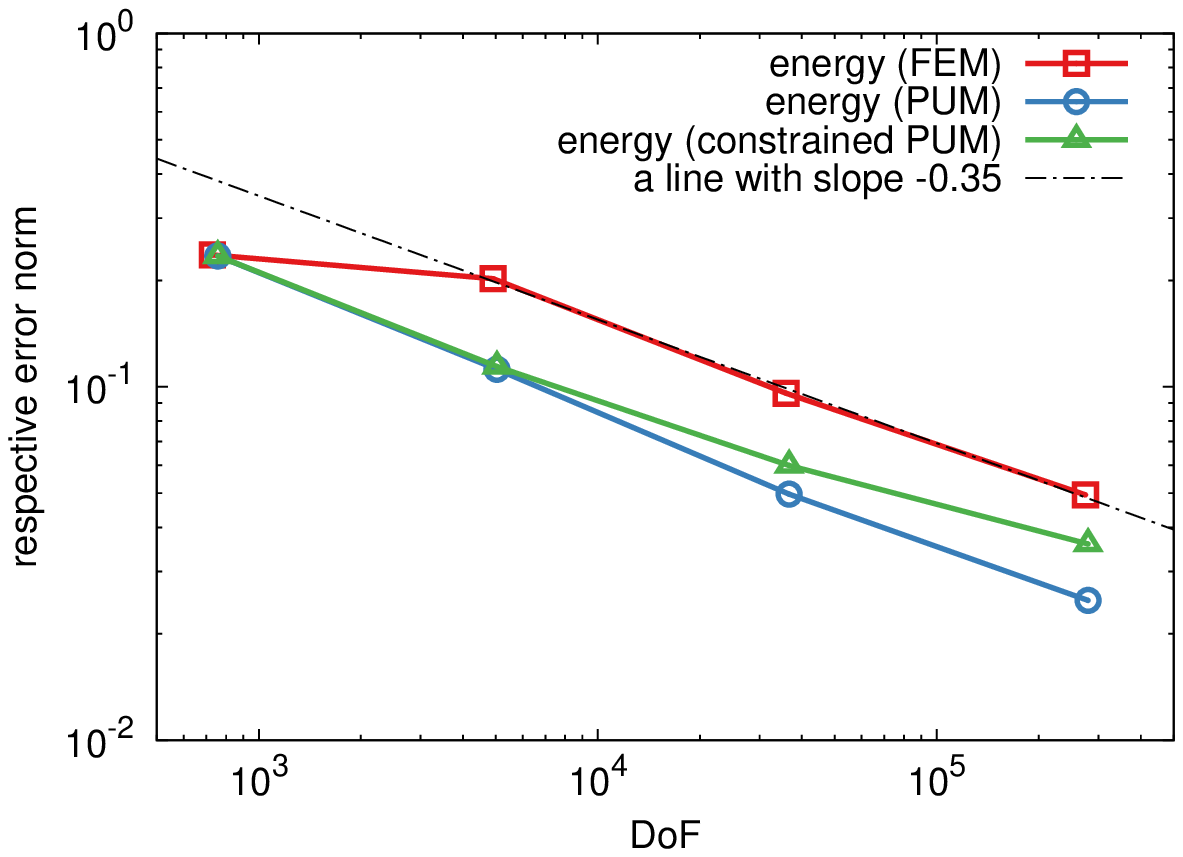}
        \caption{ 
            global.
        }%
        \label{fig:Poiss_HG_g}
    \end{subfigure}
    \caption{Convergence of the errors for the Poisson problem with $\sigma=1.0$.
    }
    \label{fig:Poiss_HG}
\end{figure}

As was mentioned in the introduction, 
a similar case was considered in \cite{Sukumar:2009vw} (albeit for periodic boundary conditions with global refinement only), 
however the authors constrained all enriched DoFs to be of the same value.
The resulting space is, obviously,  smaller than the unconstrained PUM space and thus the Galerkin projection
will certainly lead to higher errors. Figure \ref{fig:Poiss_HG_g} compares the energy error norm for the standard FEM, and constrained and unconstrained PUM in the course of global refinement for the case $\sigma=1.0$.
It is seen that by constraining PUM DoFs to have the same value, the accuracy is reduced by half for the finest mesh. 

\begin{figure}[!ht]
    \begin{subfigure}[b]{0.48\textwidth}
        \centering
        \psfrag{DoF}[t][t][0.8]{$\rm{DoFs}$}        
        \includegraphics[width=0.98\textwidth]{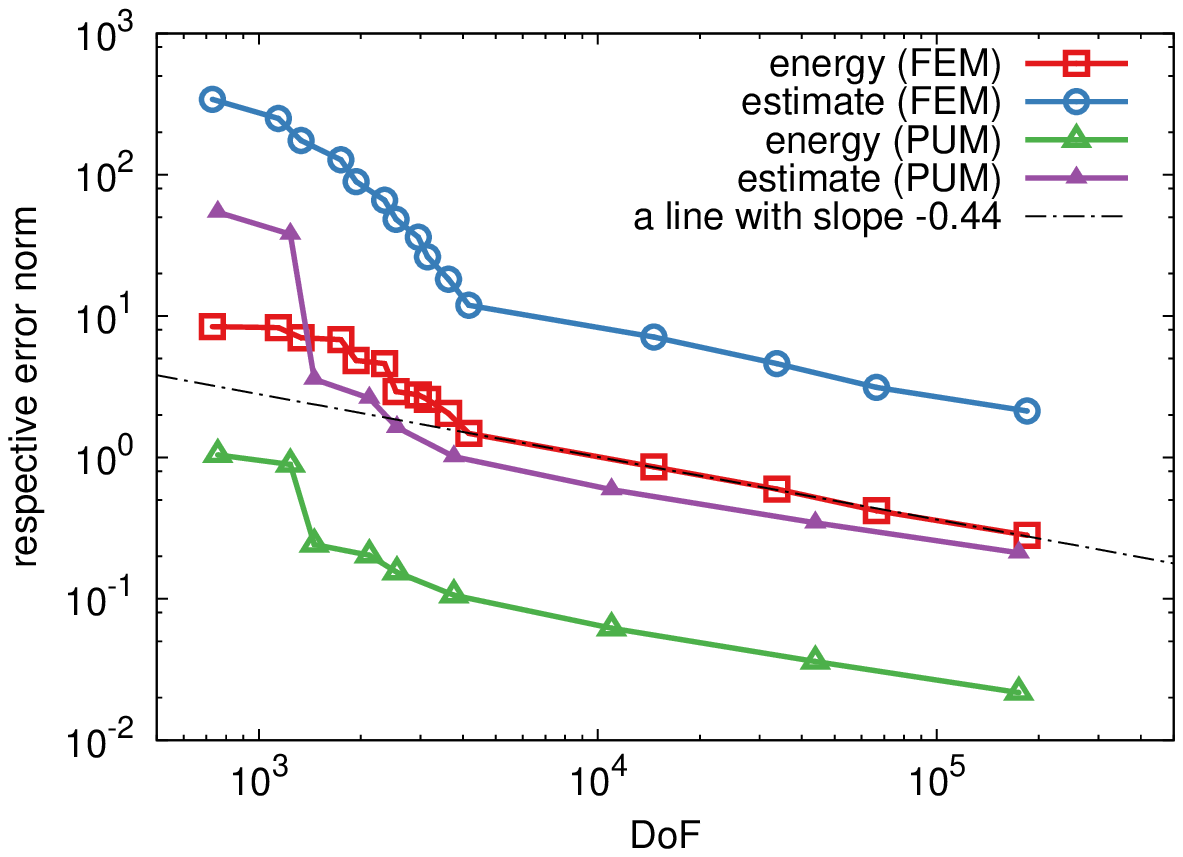}
        \caption{ 
            $h$\,-adaptive.
        }%
        \label{fig:Poiss_HG2_h}
    \end{subfigure}
    ~
    \begin{subfigure}[b]{0.48\textwidth}
        \centering
        \psfrag{DoF}[t][t][0.8]{$\rm{DoFs}$}
        \includegraphics[width=0.98\textwidth]{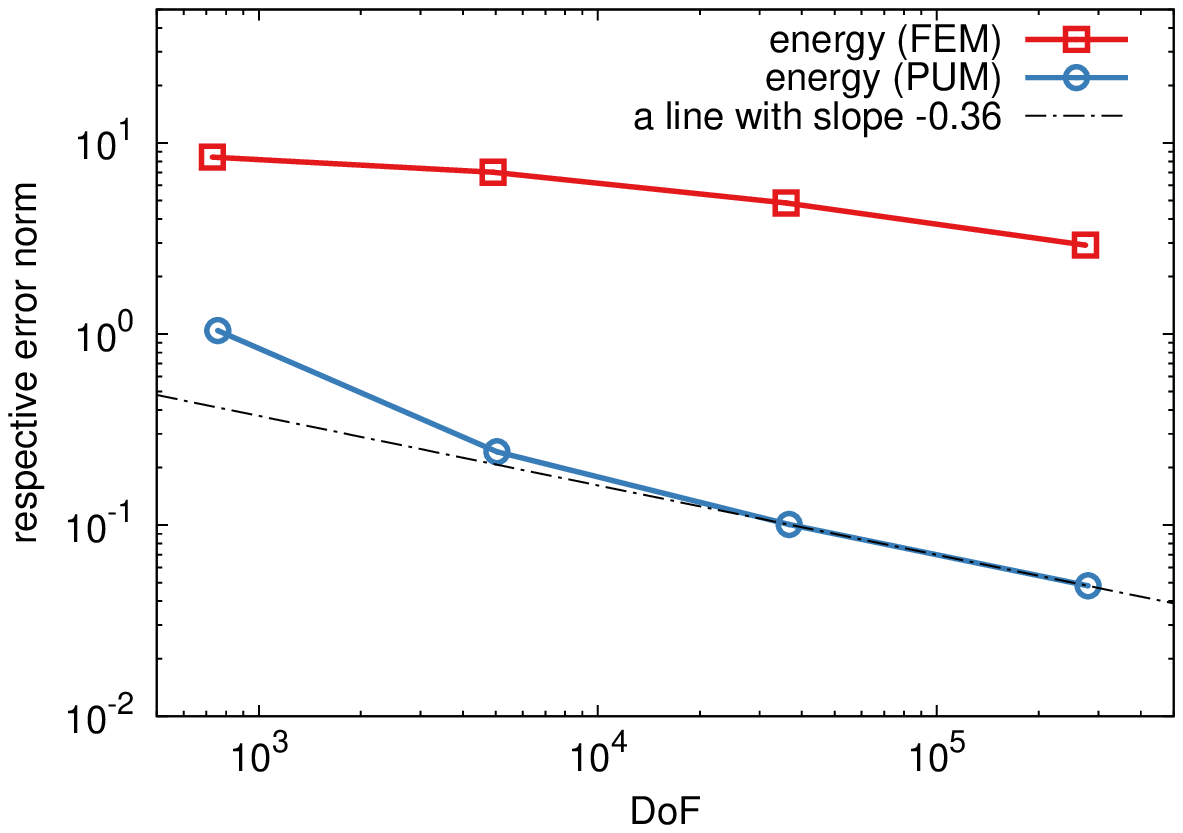}
        \caption{ 
            global.
        }%
        \label{fig:Poiss_HG2_g}
    \end{subfigure}
    \caption{Convergence of the errors for the Poisson problem with $\sigma=0.1$.
    }
    \label{fig:Poiss_HG2}
\end{figure}

Remarkably for the case $\sigma=1.0$ the PUM is only slightly more accurate than the standard FEM. The same observation can be made for the $h$\,-adaptive refinement, shown in Figure  \ref{fig:Poiss_HG_h}. 
By comparison, for the case $\sigma=0.1$ the PUM is significantly more accurate than the standard FEM, both for the case of global and $h$\,-adaptive refinement (see Figure  \ref{fig:Poiss_HG2}).
This indicates that, not surprisingly, the efficiency of the PUM as compared to FEM is very much contingent upon the underlying exact analytical solution.

Finally, we observe that the convergence rates in the case of $h$\,-adaptive refinement are roughly the same for both the standard FEM and PUM. 
This agrees with our observation for the eigenvalue problem.
Moreover, the standard residual error indicator used for the Poisson problem with PUM shows similar convergence rate for both values of $\sigma$ and therefore can be considered as a reliable error indicator for the here considered problem.

\section{Summary}
\label{sec:summary}

In this contribution we have applied the $h$\,- and $hp$\,-adaptive FEM, and the $h$\,-adaptive PUM to the relevant PDEs in quantum mechanics, namely the Schr{\"o}dinger equation and the Poisson equation. The main findings are summarized below.

\begin{itemize}
    \item The PUM renders several orders of magnitude more accurate eigenvalues than the standard FEM when solving the Schr{\"o}dinger equation for the lowest eigenpair with Coulomb and harmonic potential. For the case when more eigenpairs are sought but only the lowest eigenvector is introduced as an enrichment, the PUM is still more accurate, especially for the lowest eigenvalue. 
    Remarkably 
    other eigenvalues also exhibit a faster convergence.
    \item 
    For the here considered smoothness and residual error estimators, an application of the $hp$\,-adaptive FEM to the Hydrogen atom displays an exponential-like convergence rate for the first eigenvalue, whereas other eigenvalues tend to stagnate.
    This illustrates the challenge of applying the $hp$\,-adaptive FEM to eigenvalue problems, namely that there are multiple solution fields represented on the same FE space that are likely to have distinct smooth and non-smooth (singular) regions.  
    \item Constraining PUM DoFs to have the same value when solving the Poisson equation could decrease the accuracy of the solution by a factor of two.
    \item The efficiency of the PUM problem is very much dependent on the underlying solution. 
    On the one hand
    when applied to the Poisson problem with the here studied density field, which is composed of the Gaussian charge and the charge of the electron in the Hydrogen atom, the PUM is only slightly more accurate for the case of $\sigma=1.0$. On the other hand, for $\sigma=0.1$ the PUM is about two orders of magnitude more accurate than the standard FEM.
    \item The residual error estimator used for the Poisson problem with PUM shows a similar convergence rate to the energy error and, therefore, can be considered as a reliable error indicator for here considered problem.
    \item An element view to the implementation of PUM in FEM codes based on hexahedra is proposed. As a result, continuity of the enriched field along the edges with hanging nodes is enforced by treating FE spaces produced by each function in the local approximation space separately. The resulting algebraic constraints are independent on the enrichment functions. This allows one to directly reuse algorithms written for enforcing continuity of vector-valued FE spaces constructed from a list of scalar-valued FEs.
    \item Local interpolation error estimates are derived for the PUM enriched with the class of exponential functions. In this case the results are the same as for the standard FEM and thereby admit the usage of the error indicator (\ref{eq:residual_indicator}).
\end{itemize}

\section{Appendix: Local interpolation error estimates}
In this appendix, 
the local interpolation error estimates required for the derivation of the error indicator (\ref{eq:residual_indicator}) in the case of PUM are obtained for linear finite element approximations enriched with $ f(\gz x) = \exp{(-\mu\,\norm{\gz x}^p)}$, where $0 < \mu \in \mathbb{R}$ and $1 \leq p \in \mathbb{N}$.
These are
\begin{align}
\left\|v-q^h v\right\|_{L^2(K)}\leq c_K h_K \left|v\right|_{H^1(\omega_K)},
\label{EstimK}
\end{align}
\begin{align}
\left\|v-q^h v\right\|_{L^2(e)}\leq c_e h_K^\frac{1}{2} \left|v\right|_{H^1(\omega_K)},
\label{EstimE}
\end{align}
where, as usual, $v:\Omega\rightarrow\mathbb{R}$ is a scalar-valued function, which is assumed to be at least in $H^1(\Omega)$, $q^h$ is a quasi-interpolation operator (of the averaging type), $K$ is an element of the discretization $\mathcal{P}^h$ of $\Omega$, $e\subset\partial K$ is an edge of $K$. Also, $h_K$ measures the size of $K$, $\omega_K$ is the patch of elements neighboring $K$ including $K$ itself. Finally, $c_K,c_e\in\mathbb{R}$ are the interpolation constants independent of the mesh size.

We fix the notations to be used throughout the appendix and make assumptions that are conventional for this kind of analysis. For the sake of simplicity and without loss of generality, we elaborate here for the two-dimensional setting. The obtained results are valid in three dimensions as well.

First, we assume that the partition $\mathcal{P}^h$ of $\Omega\subset\mathbb{R}^2$ consisting of open and convex quadrilaterals $K$ is {\em shape-regular} (or {\em non-degenerate}), as well as locally {\em quasi-uniform} in the sense of \cite{Ciarl,Ming}. For every $K$ and its edge $e$ we define $h_K:=\mathrm{diam}(K)$ and $h_e:=|e|$ is the length of $e$. For every node $i$ in $\mathcal{P}^h$ we denote by $\omega_i$ the union of quadrilaterals connected to node $i$ and set $h_{\omega_i}:=\mathrm{diam}(\omega_i)$. Furthermore, for every $K$, $\omega_K$ represents the patch containing $K$ and the first row of its neighbors; it is then set $h_{\omega_K}:=\mathrm{diam}(\omega_K)$. 

Also, in what follows, by the notation $a\lesssim b$ we imply the existence of a positive constant $C$ independent of $a$ and $b$ such that $a\leq Cb$. Then $a\sim b$ means that $a\lesssim b$ and $a\gtrsim b$ hold simultaneously. The symbol $|\cdot|$ will be used to denote either the $H^1$-seminorm (as e.g. in (\ref{EstimK}) and (\ref{EstimE})) or the {\em length} of a linear segment in $\mathbb{R}^2$ or the {\em area} of a plane domain in $\mathbb{R}^2$. With these notations at hand, one can show that $|K|^\frac{1}{2}\sim h_K$, $|\omega_i|^\frac{1}{2}\sim h_{\omega_i}$ and $|\omega_K|^\frac{1}{2}\sim h_{\omega_K}$. Furthermore, the shape regularity of the mesh $\mathcal{P}^h$ ensures that $h_e\sim h_K$, whereas its local quasi-uniformity implies that $h_K\sim h_{\omega_i}\sim h_{\omega_K}$.

Finally, we also recall useful inequalities, which are 
\begin{itemize}
    \item 
the Poincar\'e-type inequality (see e.g. \cite{Veeser}):
\begin{align}
\left\|v-\frac{1}{|\omega|}\int_\omega v\; d{\gz x}\right\|_{L^2(\omega)}\leq \frac{h_\omega}{\pi} \left|v\right|_{H^1(\omega)},
\quad \forall v\in H^1(\omega),
\label{Poinc}
\end{align}
where $\omega\subset\mathbb{R}^n$ ($n=2,3$) is a Lipschitz domain and $h_\omega:=\mathrm{diam}(\omega)$;
 \item the scaled trace inequality (e.g. in \cite{Verf}, Lemma 3.2):
\begin{align}
\left\|v\right\|_{L^2(e)} \lesssim
h_e^{-\frac{1}{2}}\left\|v\right\|_{L^2(K)}+h_e^\frac{1}{2}\left|v\right|_{H^1(K)},
\quad \forall v\in H^1(K).
\label{Trace}
\end{align}
\end{itemize}

\subsection{Quasi-interpolation operator}
Herein, we construct an interpolation operator for obtaining the local error estimates (\ref{EstimK}) and (\ref{EstimE}). 

Let $V:=H^1(\Omega)$ be an admissible space and $V^h$ be its (enriched) finite element counterpart
\begin{align}
\begin{split}
V^h:=&\left\{ v^h\in C(\Omega): v^h({\gz x}):=\sum_{i\in I^\star} a_i N_i({\gz x})
+ f({\gz x})\sum_{i\in I^\star}
b_i N_i({\gz x}) \right. \\
&\left. + \sum_{i\in I^\mathrm{std.}} c_i N_i({\gz x}), \;\;a_i,b_i,c_i\in\mathbb{R} \right\} \subset V,
\end{split}
\label{Vhspace}
\end{align}
where 
$I^\star$ is the set of all enriched nodes of $\mathcal{P}^h$ and $I^\mathrm{std.}$ is the set of standard, i.e. non-enriched nodes of $\mathcal{P}^h$; $I^\star \cap I^\mathrm{strd.} = \emptyset$. Recall also that $ N_i$ in our case is the $Q_1$-shape function associated with node $i$ and supported on $\omega_i$.

\begin{figure}[!ht]
\centering
\includegraphics[width=0.5\textwidth]{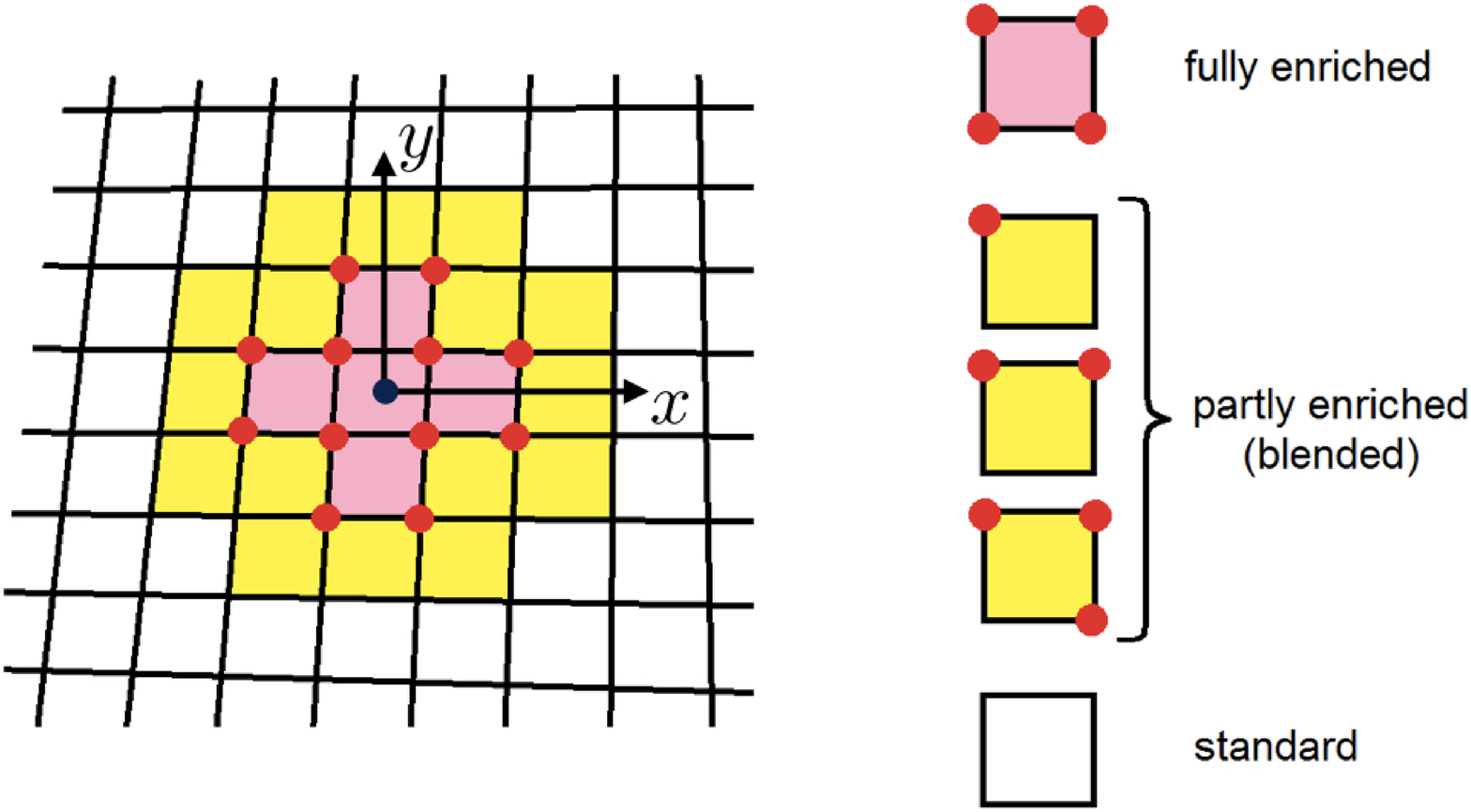}
\caption{Types of elements in mesh $\mathcal{P}^h$ with respect to the imposed enrichment.}
\label{Fig1}
\end{figure}

Explicit construction of the operator $q^h:V\rightarrow V^h$ implies the explicit pattern of assignments of $a_i,b_i,c_i\in\mathbb{R}$ through a function $v\in V$. In the case of the enriched FE approximation (\ref{Vhspace}), the major challenge in deriving $q^h$ is imposition of the constant-preserving property on $q^h$, which should be fulfilled on every element $K\in\mathcal{P}^h$ regardless the element type (see Figure \ref{Fig1}). 

The operator $q^h:V\rightarrow V^h$ with the desired property reads as follows:
\begin{align}
\begin{split}
q^h v({\gz x}):=
&\sum_{i\in I^\star} 
\left[ \frac{1}{2|\omega_i|}\int_{\omega_i} v({\gz y}) d{\gz y}\right] N_i({\gz x})
+ f({\gz x})\sum_{i\in I^\star}
\left[ \frac{1}{2f({\gz x}_i)|\omega_i|}\int_{\omega_i} v({\gz y}) d{\gz y}\right] N_i({\gz x}) \\
+&\sum_{i\in I^\mathrm{strd.}}
\left[ \frac{1}{|\omega_i|}\int_{\omega_i} v({\gz y}) d{\gz y}\right] N_i({\gz x}),
\end{split}
\label{qh}
\end{align}
with all notations as in (\ref{Vhspace}) and where ${\gz x}_i$, entering the second term, denotes the coordinate of a node $i$. 
Below, for the proposed quasi-interpolation operator of the averaging type $q^h$ we establish that $q^h c|{_K}=c$ on a standard element (note this is a classical result for a non-enriched FEM) and, more importantly, that $q^h c|{_K}= c+\mcl O(h_K^p)$ on a fully-enriched and a blended element.

\subsection{Estimates}

\subsubsection{Preliminaries.}

The three estimates that we start with are basic for the following local interpolation error analysis. On every $K\in\mathcal{P}^k$ and its node $i$ it holds that
\begin{align}
\left\|  N_i({\gz x}) \right\|_{L^2(K)} \lesssim h_K,
\quad
\left\|  N_i({\gz x}) \right\|_{L^2(e)} \lesssim h_K^\frac{1}{2},
\label{phi}
\end{align}

\begin{align}
\left| \frac{1}{|\omega_i|}\int_{\omega_i} v({\gz y}) d{\gz y} \right|
\lesssim h_K^{-1}\left\| v \right\|_{L^2(\omega_K)}+\left|v\right|_{H^1(\omega_K)},
\label{averng}
\end{align}
and
\begin{align}
\frac{f({\gz x})}{f({\gz x}_i)}= 1+\mcl O(h_K^p).
\label{Eratio}
\end{align}

Results (\ref{phi}) rigorously follow from the isoparametric concept and related properties, see e.g. \cite{Ainsworth1997} for details. We note that they may be also derived in a less rigorous manner owing to a boundedness of the basis function $ N_i$ on $K$ along with $|K|^\frac{1}{2}\sim h_K$ and $|e|^\frac{1}{2}=h_e^\frac{1}{2}\sim h_K^\frac{1}{2}$. 

The inequality (\ref{averng}) is obtained as follows:
\begin{align*}
\left| \frac{1}{|\omega_i|}\int_{\omega_i} v({\gz y}) d{\gz y} \right| 
&\leq 
|\omega_i|^{-1} \int_{\omega_i} \left| v({\gz y}) \right| d{\gz y} \leq 
|\omega_i|^{-\frac{1}{2}} \left\| v \right\|_{L^2(\omega_i)} \\
&\lesssim 
h_K^{-1}\left\| v \right\|_{L^2(\omega_K)}+\left|v\right|_{H^1(\omega_K)}.
\end{align*}
Here we used the Cauchy-Schwarz inequality, $|\omega_i|^\frac{1}{2}\sim h_{\omega_i} \sim h_K$ and also the extension-related result $\left\| v \right\|_{L^2(\omega_i)}\leq \left\| v \right\|_{L^2(\omega_K)}$. 

Finally, to show (\ref{Eratio}) we explicitly use the properties of $f(\gz x)$. For any fixed $K$, ${\gz x}\in K$ and ${\gz x}_i\in K$ being one of its nodes, we have the following upper bound estimate:
\begin{align}
\begin{split}
\frac{f({\gz x})}{f({\gz x}_i)} 
&= \frac{\exp(\mu|\gz x_i|^p)}{\exp(\mu|\gz x|^p)}
\leq \frac
{\exp\left(\mu\left[\max_{\gz x\in\overline{K}}|\gz x|\right]^p \right)}
{\exp\left(\mu\left[\min_{\gz x\in\overline{K}}|\gz x|\right]^p \right)} 
\leq \frac
{\exp\left( \mu\left[\min_{\gz x\in\overline{K}}|\gz x|+h_K\right]^p \right)}
{\exp\left( \mu\left[\min_{\gz x\in\overline{K}}|\gz x|\right]^p \right)} 
 \\
& = 1+\mu p\left[ \min_{\gz x\in\overline{K}}|\gz x| \right]^{p-1}h_K
+ \mathrm{h.o.t.\; in\;} \left\{ \min_{\gz x\in\overline{K}}|\gz x|,h_K \right\}
 \\
& = 1+\mcl O\left(\left[ \min_{\gz x\in\overline{K}}|\gz x| \right]^{p-1}h_K \right).
\end{split}
\label{New1}
\end{align}
Notice that due to boundedness of $\min_{\gz x\in\overline{K}}|\gz x|$ for a given fixed $K$, there always exists $\epsilon>0$ such that $\min_{\gz x\in\overline{K}}|\gz x|=\epsilon h_K$. Using this in (\ref{New1}), we obtain
\begin{align*}
\left[\min_{\gz x\in\overline{K}}|\gz x| \right]^{p-1}h_K = \epsilon^{p-1}h_K^p,
\end{align*}
yielding, as a result, $\frac{f({\gz x})}{f({\gz x}_i)}\leq 1+\mcl O(h_K^p)$.

The lower bound estimate can be found similarly:
\begin{align*}
\frac{f({\gz x})}{f({\gz x}_i)} 
&= \frac{\exp(\mu|\gz x_i|^p)}{\exp(\mu|\gz x|^p)}
\geq \frac
{\exp\left( \mu\left[\min_{\gz x\in\overline{K}}|\gz x|\right]^p \right)}
{\exp\left( \mu\left[\max_{\gz x\in\overline{K}}|\gz x|\right]^p \right)} 
\geq \frac
{\exp\left( \mu\left[\min_{\gz x\in\overline{K}}|\gz x|\right]^p \right)}
{\exp\left( \mu\left[\min_{\gz x\in\overline{K}}|\gz x|+h_K\right]^p \right)} \\
& = 1-\mu p\left[ \min_{\gz x\in\overline{K}}|\gz x| \right]^{p-1}h_K
+ \mathrm{h.o.t.\; in\;} \left\{ \min_{\gz x\in\overline{K}}|\gz x|,h_K \right\} \\
& = 1+\mcl O\left(\left[ \min_{\gz x\in\overline{K}}|\gz x| \right]^{p-1}h_K \right),
\end{align*}
and, eventually, $\frac{f({\gz x})}{f({\gz x}_i)}\geq 1+\mcl O(h_K^p)$. The result (\ref{Eratio}) then follows.

\subsubsection{Stability of $q^h$ in $L^2$-norm.}

The next step towards (\ref{EstimK}) and (\ref{EstimE}) implies obtaining the so-called stability result for the constructed $q^h$. Using (\ref{phi})--(\ref{Eratio}) one straightforwardly shows that
\begin{align}
\left\| q^h v \right\|_{L^2(K)}
\lesssim 
\left\| v \right\|_{L^2(\omega_K)}+h_K\left|v\right|_{H^1(\omega_K)},
\label{stab1}
\end{align}
and
\begin{align}
\left\| q^h v \right\|_{L^2(e)}
\lesssim 
h_K^{-\frac{1}{2}} \left\| v \right\|_{L^2(\omega_K)} + h_K^\frac{1}{2} \left|v\right|_{H^1(\omega_K)}.
\label{stab2}
\end{align}
These estimates indeed hold for every $K$ regardless of its type (standard, blended, enriched). Note that for a standard non-enriched FEM and the resulting interpolation operators, the estimates (\ref{stab1}), (\ref{stab2}) are classical. We have obtained and proved them for our specific operator $q^h$ adopted for the current enriched FEM setting.

\subsubsection{Constant-preserving property of $q^h$.}
\label{sec:constant-preserving-prop}

The final ingredient required for obtaining (\ref{EstimK}) and (\ref{EstimE}) is the determination of how ``well" the constructed $q^h$ reproduces the constant on an element $K$, depending on its type. This constant-preserving property of the operator is of major importance particularly in the case of enriched FEM.

The required result on a standard (non-enriched) element $K$ follows immediately. Indeed, in this case
\begin{align*}
q^h v({\gz x})|_K=
\sum_{i=1}^{4}
\left[ \frac{1}{|\omega_i|}\int_{\omega_i} v({\gz y}) d{\gz y}\right] N_i({\gz x}),
\end{align*}
and the partition of unity $\sum_{i=1}^{4} N_i({\gz x})=1$ on $K$ yields $q^h c|_K=c$, $c=\mathrm{const}$.

The situation on a fully-enriched and partly-enriched (blended) element is more delicate.
In the case of a fully enriched element we have
\begin{align*}
q^h v({\gz x})|_K=
&\sum_{i=1}^{4}
\left[ \frac{1}{2|\omega_i|}\int_{\omega_i} v({\gz y}) d{\gz y}\right] N_i({\gz x}) \\
+ 
&f({\gz x})\sum_{i=1}^{4}
\left[ \frac{1}{2f({\gz x}_i)|\omega_i|}\int_{\omega_i} v({\gz y}) d{\gz y}\right] N_i({\gz x}),
\end{align*}
that, owing to (\ref{Eratio}), results in
\begin{align*}
q^h c|_K=
\frac{1}{2}c + \frac{1}{2}c
\sum_{i=1}^{4} \frac{f({\gz x})}{f({\gz x}_i)} N_i({\gz x})
= 
\frac{1}{2}c + \frac{1}{2}c\left[1+\mcl O(h_K^p)\right]\sum_{i=1}^{4} N_i({\gz x})
=c+\mcl O(h_K^p).
\end{align*}

Now, let $K$ be a blended element, implying the representation:
\begin{align*}
q^h v({\gz x})|_K&=
\sum_{i=1}^{\ell}
\left[ \frac{1}{2|\omega_i|}\int_{\omega_i} v({\gz y}) d{\gz y}\right] N_i({\gz x}) \\
&+ f({\gz x})\sum_{i=1}^{\ell}
\left[ \frac{1}{2f({\gz x}_i)|\omega_i|}\int_{\omega_i} v({\gz y}) d{\gz y}\right] N_i({\gz x}) \\
&+\sum_{i=\ell+1}^{4}
\left[ \frac{1}{|\omega_i|}\int_{\omega_i} v({\gz y}) d{\gz y}\right] N_i({\gz x}),
\end{align*}
where $\ell\in\{1,2,3\}$ is the number of enriched nodes of $K$. Adding and subtracting the first sum in the above expression, enables us to rewrite it as follows:
\begin{align*}
q^h v({\gz x})|_K&=
-\sum_{i=1}^{\ell}
\left[ \frac{1}{2|\omega_i|}\int_{\omega_i} v({\gz y}) d{\gz y}\right] N_i({\gz x}) \\
&+ f({\gz x})\sum_{i=1}^{\ell}
\left[ \frac{1}{2f({\gz x}_i)|\omega_i|}\int_{\omega_i} v({\gz y}) d{\gz y}\right] N_i({\gz x})\\
&+\sum_{i=1}^{4}
\left[ \frac{1}{|\omega_i|}\int_{\omega_i} v({\gz y}) d{\gz y}\right] N_i({\gz x}).
\end{align*}
Note that the last term contains the summation over all four nodes and is the standard (non-enriched) FE contribution which will automatically reproduce a constant. We then need to estimate, in this context, the remaining part constituting of the first and the second sums. We obtain,
\begin{align*}
q^h c|_K
&=\frac{1}{2}\,c \sum_{i=1}^{\ell}
\left[ \frac{f({\gz x})}{f({\gz x}_i)}-1\right] N_i({\gz x}) +c
\leq
\frac{1}{2}\,c \sum_{i=1}^{\ell}
\norm{ \frac{f({\gz x})}{f({\gz x}_i)}-1} N_i({\gz x}) +c \\
&\leq
\frac{1}{2}\,c \sum_{i=1}^{4}
\norm{\frac{f({\gz x})}{f({\gz x}_i)}-1} N_i({\gz x}) +c
=
\frac{1}{2}\,c \,\mcl O(h_K^p) \sum_{i=1}^{4} N_i({\gz x}) +c
=c+\mcl O(h_K^p),
\end{align*}
where (\ref{Eratio}) was also used.

\subsubsection{Proof of local error estimates (\ref{EstimK}), (\ref{EstimE})}

The derivation of the estimates for $\left\|v-q^h v\right\|_{L^2(K)}$ and $\left\|v-q^h v\right\|_{L^2(e)}$ is based on a combined use of the above stability results for $q^h$, the Poincar\'e and the scaled trace inequalities (\ref{Poinc}) and (\ref{Trace}), respectively, as well as the constant-preserving property results. First, due to linearity of $q^h$, we have
\begin{align}
\begin{split}
\left\| v - q^h v \right\|_{L^2(\sigma)} &=
\left\| v - c - q^h(v-c) + c-q^h c  \right\|_{L^2(\sigma)} \\
&\leq 
\left\| v - c \right\|_{L^2(\sigma)}
+ \left\| q^h(v-c) \right\|_{L^2(\sigma)}
+ \left\| c-q^h c  \right\|_{L^2(\sigma)},
\end{split}
\label{interp}
\end{align}
where $c=\mathrm{const}$ and where, for the sake of brevity, we set $\sigma=\{K,e\}$. We are now in a position to dissect every term in (\ref{interp}) in either case of $\sigma$.\\

{\bf When $\sigma=K$ in (\ref{interp}):} \\

By the Poincar\'e inequality (\ref{Poinc}), it holds that
\begin{align}
\left\| v - c \right\|_{L^2(K)} 
\leq \left\| v - c \right\|_{L^2(\omega_K)} 
\lesssim h_K \left|v\right|_{H^1(\omega_K)},
\label{1a}
\end{align}
where one can choose $c=|\omega_K|^{-1}\int_{\omega_K} v \rm d{\gz x}$ and use $h_{\omega_K}\sim h_K$.

By the stability estimate (\ref{stab1}) and the Poincar\'e inequality, it holds similarly to the above that
\begin{align}
\left\| q^h(v-c) \right\|_{L^2(K)} 
\lesssim \left\| v-c \right\|_{L^2(\omega_K)} +h_K \left| v-c \right|_{H^1(\omega_K)}
\lesssim h_K \left|v\right|_{H^1(\omega_K)}.
\label{1b}
\end{align}

Furthermore, using the results of Section \ref{sec:constant-preserving-prop} we obtain
\begin{align}
\left\| c-q^h c  \right\|_{L^2(K)}\equiv0, \quad \mbox{if}\; K\; \mbox{is standard},
\label{1c}
\end{align}
and 
\begin{align}
\left\| c-q^h c  \right\|_{L^2(K)}= \mcl O(h_K^{p+1}), \quad \mbox{if}\; K\; \mbox{is fully enriched or blended}.
\label{1d}
\end{align}
In the former case we also use that $\left\| 1  \right\|_{L^2(K)}=|K|^\frac{1}{2}\sim h_K$. 

Using (\ref{1a})--(\ref{1d}) in (\ref{interp}), the resulting local interpolation error of type (\ref{EstimK}) follows. Note that in the case of fully enriched and blended elements the term $\mcl O(h_K^{p+1})$ that appears in the corresponding upper bound can be neglected, being the higher order term with respect to the leading one $h_K \left|v\right|_{H^1(\omega_K)}$.\\

{\bf When $\sigma=e$ in (\ref{interp}):} \\

By the scaled trace inequality (\ref{Trace}), it holds
\begin{align}
\left\| v - c \right\|_{L^2(e)} 
\lesssim h_e^{-\frac{1}{2}}\left\|v-c\right\|_{L^2(K)}+h_e^\frac{1}{2}\left|v-c\right|_{H^1(K)}
\lesssim h_K^\frac{1}{2} \left|v\right|_{H^1(\omega_K)},
\label{2a}
\end{align}
where we also use $h_e\sim h_K$ along with result in (\ref{1a}).

By the stability estimate (\ref{stab2}) and the Poincar\'e inequality (\ref{Poinc}), we obtain the result that
\begin{align}
\left\| q^h(v-c) \right\|_{L^2(e)} 
\lesssim h_K^{-\frac{1}{2}} \left\| v-c \right\|_{L^2(\omega_K)} +h_K^\frac{1}{2} \left| v-c \right|_{H^1(\omega_K)}
\lesssim h_K^\frac{1}{2} \left|v\right|_{H^1(\omega_K)}.
\label{2b}
\end{align}

Finally, using the results of section \ref{sec:constant-preserving-prop} we derive
\begin{align}
\left\| c-q^h c  \right\|_{L^2(e)}\equiv0, \quad \mbox{if}\; K\; \mbox{is standard},
\label{2c}
\end{align}
and 
\begin{align}
\left\| c-q^h c  \right\|_{L^2(K)}= \mcl O(h_K^{p+\frac{1}{2}}), \quad \mbox{if}\; K\; \mbox{is fully enriched or blended}.
\label{2d}
\end{align}
In the former case we also use the fact that $\left\| 1  \right\|_{L^2(e)}=|e|^\frac{1}{2}=h_e^\frac{1}{2}\sim h_K^\frac{1}{2}$. \\

Using (\ref{2a})--(\ref{2d}) in (\ref{interp}), the resulting local interpolation error estimate of type (\ref{EstimE}) follows as well. Again, in the case of fully enriched and blended elements the term $\mcl O(h_K^\frac{3}{2})$ that appears in the corresponding upper bound can be neglected, being the higher order term with respect to the leading one $h_K^\frac{1}{2} \left|v\right|_{H^1(\omega_K)}$.

\section*{Acknowledgements}

The support of this work by the ERC Advanced Grant 289049 MOCOPOLY (DD,JP,PS) and
the Competence Network for Technical and Scientific High Performance Computing in Bavaria (KONWIHR) (DD) is gratefully acknowledged.
Second author (TG) is supported by the European Research Council (ERC) Starting Researcher Grant INTERFACES, Grant Agreement N. 279439.

\bibliographystyle{abbrvnat}
\bibliography{Bibliography}

\begin{thebibliography}{60}
\providecommand{\natexlab}[1]{#1}
\providecommand{\url}[1]{\texttt{#1}}
\expandafter\ifx\csname urlstyle\endcsname\relax
  \providecommand{\doi}[1]{doi: #1}\else
  \providecommand{\doi}{doi: \begingroup \urlstyle{rm}\Url}\fi

\bibitem[Ainsworth and Oden(1997)]{Ainsworth1997}
M.~Ainsworth and J.~T. Oden.
\newblock A posteriori error estimation in finite element analysis.
\newblock \emph{Comput. Methods Appl. Mech. Engrg.}, 142:\penalty0 1--88, 1997.

\bibitem[Babu\v{s}ka and Melenk(1997)]{Babuska1997}
I.~Babu\v{s}ka and J.~M. Melenk.
\newblock The partition of unity method.
\newblock \emph{International Journal for Numerical Methods in Engineering},
  40\penalty0 (4):\penalty0 727--758, 1997.
\newblock ISSN 1097-0207.
\newblock
  \doi{10.1002/(SICI)1097-0207(19970228)40:4<727::AID-NME86>3.0.CO;2-N}.

\bibitem[Balay et~al.(2015)Balay, Abhyankar, Adams, Brown, Brune, Buschelman,
  Dalcin, Eijkhout, Gropp, Kaushik, Knepley, McInnes, Rupp, Smith, Zampini, and
  Zhang]{petsc-user-ref}
S.~Balay, S.~Abhyankar, M.~F. Adams, J.~Brown, P.~Brune, K.~Buschelman,
  L.~Dalcin, V.~Eijkhout, W.~D. Gropp, D.~Kaushik, M.~G. Knepley, L.~C.
  McInnes, K.~Rupp, B.~F. Smith, S.~Zampini, and H.~Zhang.
\newblock {PETS}c users manual.
\newblock Technical Report ANL-95/11 - Revision 3.6, Argonne National
  Laboratory, 2015.
\newblock URL \url{http://www.mcs.anl.gov/petsc}.

\bibitem[Bangerth()]{dealii-tutorial27}
W.~Bangerth.
\newblock The \texttt{deal.II} library tutorial step 27 (version 8.3).
\newblock URL \url{https://www.dealii.org/8.3.0/doxygen/deal.II/step_27.html}.
\newblock Accessed on January 2016.

\bibitem[Bangerth and Kayser-herold(2009)]{Bangerth:2008uw}
W.~Bangerth and O.~Kayser-herold.
\newblock {Data Structures and Requirements for hp Finite Element Software}.
\newblock \emph{ACM Transactions on Mathematical Software}, 36\penalty0
  (1):\penalty0 4, Aug. 2009.

\bibitem[Bangerth et~al.(2016)Bangerth, Davydov, Heister, Heltai, Kanschat,
  Kronbichler, Maier, Turcksin, and Wells]{dealII84}
W.~Bangerth, D.~Davydov, T.~Heister, L.~Heltai, G.~Kanschat, M.~Kronbichler,
  M.~Maier, B.~Turcksin, and D.~Wells.
\newblock The \texttt{deal.II} library, version 8.4.
\newblock \emph{Archive of Numerical Software}, 4\penalty0 (100):\penalty0
  1--11, 2016.
\newblock ISSN 2197-8263.
\newblock \doi{10.11588/ans.2016.100.23122}.

\bibitem[Bao et~al.(2012)Bao, Hu, and Liu]{Bao:2012it}
G.~Bao, G.~Hu, and D.~Liu.
\newblock {Numerical Solution of the Kohn-Sham Equation by Finite Element
  Methods with an Adaptive Mesh Redistribution Technique}.
\newblock \emph{Journal of Scientific Computing}, 55\penalty0 (2):\penalty0
  372--391, Sept. 2012.
\newblock \doi{10.1007/s10915-012-9636-1}.

\bibitem[Belytschko and Black(1999)]{Belytschko1999}
T.~Belytschko and T.~Black.
\newblock Elastic crack growth in finite elements with minimal remeshing.
\newblock \emph{International journal for numerical methods in engineering},
  45\penalty0 (5):\penalty0 601--620, 1999.

\bibitem[Belytschko et~al.(2009)Belytschko, Gracie, and
  Ventura]{Belytschko2009}
T.~Belytschko, R.~Gracie, and G.~Ventura.
\newblock A review of extended/generalized finite element methods for material
  modeling.
\newblock \emph{Modelling and Simulation in Materials Science and Engineering},
  17\penalty0 (4):\penalty0 043001, 2009.

\bibitem[Bylaska et~al.(2009)Bylaska, Holst, and Weare]{Bylaska:2009im}
E.~J. Bylaska, M.~Holst, and J.~H. Weare.
\newblock {Adaptive Finite Element Method for Solving the Exact Kohn-Sham
  Equation of Density Functional Theory}.
\newblock \emph{Journal of Chemical Theory and Computation}, 5\penalty0
  (4):\penalty0 937--948, Apr. 2009.
\newblock \doi{10.1021/ct800350j}.

\bibitem[Chahine et~al.(2008)Chahine, Laborde, and Renard]{Chahine2008}
E.~Chahine, P.~Laborde, and Y.~Renard.
\newblock Crack tip enrichment in the xfem using a cutoff function.
\newblock \emph{International journal for numerical methods in engineering},
  75\penalty0 (6):\penalty0 629--646, 2008.

\bibitem[Ciarlet(1978)]{Ciarl}
P.~G. Ciarlet.
\newblock \emph{The finite element method for elliptic problems}.
\newblock North Holland: Amsterdam, 1978.

\bibitem[Cimrman et~al.(2015)Cimrman, Nov{\'a}k, Kolman, T{\u{u}}ma, and
  Vack{\'a}{\v{r}}]{Cimrman2015}
R.~Cimrman, M.~Nov{\'a}k, R.~Kolman, M.~T{\u{u}}ma, and J.~Vack{\'a}{\v{r}}.
\newblock Finite element method and isogeometric analysis in electronic
  structure calculations: convergence study.
\newblock \emph{arXiv preprint arXiv:1512.07156}, 2015.

\bibitem[Dai et~al.(2008)Dai, Xu, and Zhou]{Dai2008}
X.~Dai, J.~Xu, and A.~Zhou.
\newblock Convergence and optimal complexity of adaptive finite element
  eigenvalue computations.
\newblock \emph{Numerische Mathematik}, 110\penalty0 (3):\penalty0 313--355,
  2008.
\newblock ISSN 0029-599X.
\newblock \doi{10.1007/s00211-008-0169-3}.
\newblock URL \url{http://dx.doi.org/10.1007/s00211-008-0169-3}.

\bibitem[Davydov et~al.(2015)Davydov, Young, and Steinmann]{Davydov2014}
D.~Davydov, T.~Young, and P.~Steinmann.
\newblock {On the adaptive finite element analysis of the Kohn-Sham equations:
  Methods, algorithms, and implementation.}
\newblock \emph{Journal for Numerical Methods in Engineering. Accepted}, 2015.

\bibitem[Dolbow and Belytschko(1999)]{Dolbow1999}
J.~Dolbow and T.~Belytschko.
\newblock A finite element method for crack growth without remeshing.
\newblock \emph{Int. J. Numer. Meth. Eng}, 46\penalty0 (1):\penalty0 131--150,
  1999.

\bibitem[Dur{\'a}n et~al.(2003)Dur{\'a}n, Padra, and Rodr{\'\i}guez]{Duran2003}
R.~G. Dur{\'a}n, C.~Padra, and R.~Rodr{\'\i}guez.
\newblock A posteriori error estimates for the finite element approximation of
  eigenvalue problems.
\newblock \emph{Mathematical Models and Methods in Applied Sciences},
  13\penalty0 (08):\penalty0 1219--1229, 2003.
\newblock URL
  \url{http://www.worldscientific.com/doi/abs/10.1142/S0218202503002878}.

\bibitem[Eibner and Melenk(2007)]{Eibner2007}
T.~Eibner and J.~M. Melenk.
\newblock An adaptive strategy for hp-fem based on testing for analyticity.
\newblock \emph{Computational Mechanics}, 39\penalty0 (5):\penalty0 575--595,
  2007.

\bibitem[Fang et~al.(2012)Fang, Gao, and Zhou]{Fang:2012uu}
J.~Fang, X.~Gao, and A.~Zhou.
\newblock {A Kohn-Sham equation solver based on hexahedral finite elements}.
\newblock \emph{Journal of Computational Physics}, 231\penalty0 (8):\penalty0
  3166--3180, 2012.

\bibitem[Fankhauser et~al.(2014)Fankhauser, Wihler, and Wirz]{Fankhauser2014}
T.~Fankhauser, T.~P. Wihler, and M.~Wirz.
\newblock The hp-adaptive fem based on continuous sobolev embeddings: Isotropic
  refinements.
\newblock \emph{Computers \& Mathematics with Applications}, 67\penalty0
  (4):\penalty0 854--868, 2014.

\bibitem[Fattebert et~al.(2007)Fattebert, Hornung, and
  Wissink]{Fattebert:2007fy}
J.~L. Fattebert, R.~D. Hornung, and A.~M. Wissink.
\newblock {Finite element approach for density functional theory calculations
  on locally-refined meshes}.
\newblock \emph{Journal of Computational Physics}, 223\penalty0 (2):\penalty0
  759--773, May 2007.
\newblock \doi{10.1016/j.jcp.2006.10.013}.

\bibitem[Fries(2008)]{Fries2008}
T.-P. Fries.
\newblock A corrected xfem approximation without problems in blending elements.
\newblock \emph{International Journal for Numerical Methods in Engineering},
  75\penalty0 (5):\penalty0 503--532, 2008.

\bibitem[Fries and Belytschko(2010)]{Fries2010}
T.-P. Fries and T.~Belytschko.
\newblock The extended/generalized finite element method: an overview of the
  method and its applications.
\newblock \emph{International Journal for Numerical Methods in Engineering},
  84\penalty0 (3):\penalty0 253--304, 2010.

\bibitem[Garau et~al.(2009)Garau, Morin, and Zuppa]{Garau2009}
E.~M. Garau, P.~Morin, and C.~Zuppa.
\newblock Convergence of adaptive finite element methods for eigenvalue
  problems.
\newblock \emph{Mathematical Models and Methods in Applied Sciences},
  19\penalty0 (05):\penalty0 721--747, 2009.
\newblock URL
  \url{http://www.worldscientific.com/doi/pdf/10.1142/S0218202509003590}.

\bibitem[Gerasimov et~al.(2012)Gerasimov, R{\"u}ter, and Stein]{Gerasimov2012}
T.~Gerasimov, M.~R{\"u}ter, and E.~Stein.
\newblock An explicit residual-type error estimator for q1-quadrilateral
  extended finite element method in two-dimensional linear elastic fracture
  mechanics.
\newblock \emph{International Journal for Numerical Methods in Engineering},
  90\penalty0 (9):\penalty0 1118--1155, 2012.

\bibitem[Giani et~al.(2012)Giani, Grubi{\v{s}}i{\'c}, and Ovall]{Giani2012}
S.~Giani, L.~Grubi{\v{s}}i{\'c}, and J.~S. Ovall.
\newblock Benchmark results for testing adaptive finite element eigenvalue
  procedures.
\newblock \emph{Applied numerical mathematics}, 62\penalty0 (2):\penalty0
  121--140, 2012.

\bibitem[Griffiths(2005)]{Griffiths:2005tq}
D.~J. Griffiths.
\newblock \emph{{Introduction to Quantum Mechanics}}.
\newblock Pearson, 2 edition, 2005.

\bibitem[Hartmann and Houston(2010)]{Hartmann2010}
R.~Hartmann and P.~Houston.
\newblock Error estimation and adaptive mesh refinement for aerodynamic flows.
\newblock In \emph{ADIGMA-A European Initiative on the Development of Adaptive
  Higher-Order Variational Methods for Aerospace Applications}, pages 339--353.
  Springer, 2010.

\bibitem[Hernandez et~al.(2005)Hernandez, Roman, and Vidal]{Hernandez2005}
V.~Hernandez, J.~E. Roman, and V.~Vidal.
\newblock {SLEPc: A scalable and flexible toolkit for the solution of
  eigenvalue problems}.
\newblock \emph{ACM Transactions on Mathematical Software}, 31\penalty0
  (3):\penalty0 351--362, \#sep\# 2005.
\newblock \doi{10.1145/1089014.1089019}.
\newblock URL
  \url{http://portal.acm.org/citation.cfm?id=1089014.1089019&coll=DL&dl=ACM&CFID=239559931&CFTOKEN=59070976}.

\bibitem[Heroux et~al.(2005)Heroux, Bartlett, Howle, Hoekstra, Hu, Kolda,
  Lehoucq, Long, Pawlowski, Phipps, Salinger, Thornquist, Tuminaro,
  Willenbring, Williams, and Stanley]{Heroux2005}
M.~A. Heroux, R.~A. Bartlett, V.~E. Howle, R.~J. Hoekstra, J.~J. Hu, T.~G.
  Kolda, R.~B. Lehoucq, K.~R. Long, R.~P. Pawlowski, E.~T. Phipps, A.~G.
  Salinger, H.~K. Thornquist, R.~S. Tuminaro, J.~M. Willenbring, A.~Williams,
  and K.~S. Stanley.
\newblock An overview of the trilinos project.
\newblock \emph{ACM Trans. Math. Softw.}, 31\penalty0 (3):\penalty0 397--423,
  2005.
\newblock ISSN 0098-3500.
\newblock \doi{http://doi.acm.org/10.1145/1089014.1089021}.

\bibitem[Heuveline and Rannacher(2001)]{Heuveline2001}
V.~Heuveline and R.~Rannacher.
\newblock A posteriori error control for finite element approximations of
  elliptic eigenvalue problems.
\newblock \emph{Advances in Computational Mathematics}, 15\penalty0
  (1-4):\penalty0 107--138, 2001.
\newblock URL \url{http://link.springer.com/article/10.1023/A:1014291224961}.

\bibitem[Heuveline and Rannacher(2003)]{Heuveline2003}
V.~Heuveline and R.~Rannacher.
\newblock Duality-based adaptivity in the hp-finite element method.
\newblock \emph{Journal of Numerical Mathematics jnma}, 11\penalty0
  (2):\penalty0 95--113, 2003.

\bibitem[Hohenberg and Kohn(1964)]{Hohenberg:1964ut}
P.~Hohenberg and W.~Kohn.
\newblock {Inhomogeneous electron gas}.
\newblock \emph{Physical Review}, 136\penalty0 (3B):\penalty0 B864--B871, 1964.

\bibitem[Houston and S{\"u}li(2005)]{Houston2005}
P.~Houston and E.~S{\"u}li.
\newblock A note on the design of hp-adaptive finite element methods for
  elliptic partial differential equations.
\newblock \emph{Computer Methods in Applied Mechanics and Engineering},
  194\penalty0 (2):\penalty0 229--243, 2005.

\bibitem[Houston et~al.(2003)Houston, Senior, and S{\"u}li]{Houston2003}
P.~Houston, B.~Senior, and E.~S{\"u}li.
\newblock Sobolev regularity estimation for hp-adaptive finite element methods.
\newblock In F.~Brezzi, A.~Buffa, S.~Corsaro, and A.~Murli, editors,
  \emph{Numerical Mathematics and Advanced Applications}, pages 631--656.
  Springer Milan, 2003.
\newblock ISBN 978-88-470-2167-9.
\newblock \doi{10.1007/978-88-470-2089-4_58}.

\bibitem[Kohn and Sham(1965)]{Kohn:1965ui}
W.~Kohn and L.~J. Sham.
\newblock {Self-consistent equations including exchange and correlation
  effects}.
\newblock \emph{Physical Review}, 140\penalty0 (4A):\penalty0 A1133--A1138,
  1965.

\bibitem[Laborde et~al.(2005)Laborde, Pommier, Renard, and
  Sala{\"u}n]{Laborde2005}
P.~Laborde, J.~Pommier, Y.~Renard, and M.~Sala{\"u}n.
\newblock High-order extended finite element method for cracked domains.
\newblock \emph{International Journal for Numerical Methods in Engineering},
  64\penalty0 (3):\penalty0 354--381, 2005.

\bibitem[Larson(2000)]{Larson2000}
M.~G. Larson.
\newblock A posteriori and a priori error analysis for finite element
  approximations of self-adjoint elliptic eigenvalue problems.
\newblock \emph{SIAM journal on numerical analysis}, 38\penalty0 (2):\penalty0
  608--625, 2000.
\newblock URL \url{http://epubs.siam.org/doi/abs/10.1137/S0036142997320164}.

\bibitem[Linder(2012)]{Linder:2012tg}
C.~Linder.
\newblock \emph{{On the Computational Modeling of Micromechanical Phenomena in
  Solid Materials}}.
\newblock Habilitation thesis, Institut f{\"u}r Mechanick (Bauwesen) der
  Universit{\"a}t Stuttgart, 2012.

\bibitem[Maday(2014)]{Maday2014}
Y.~Maday.
\newblock hp finite element approximation for full-potential electronic
  structure calculations.
\newblock \emph{Chinese Annals of Mathematics, Series B}, 35\penalty0
  (1):\penalty0 1--24, 2014.

\bibitem[Mao et~al.(2006)Mao, Shen, and Zhou]{Mao2006}
D.~Mao, L.~Shen, and A.~Zhou.
\newblock Adaptive finite element algorithms for eigenvalue problems based on
  local averaging type a posteriori error estimates.
\newblock \emph{Advances in Computational Mathematics}, 25\penalty0
  (1-3):\penalty0 135--160, 2006.
\newblock URL \url{http://link.springer.com/article/10.1007/s10444-004-7617-0}.

\bibitem[Mavriplis(1994)]{Mavriplis1994}
C.~Mavriplis.
\newblock Adaptive mesh strategies for the spectral element method.
\newblock \emph{Computer methods in applied mechanics and engineering},
  116\penalty0 (1):\penalty0 77--86, 1994.

\bibitem[Melenk and Babu\v{s}ka(1996)]{Melenk1996}
J.~Melenk and I.~Babu\v{s}ka.
\newblock The partition of unity finite element method: Basic theory and
  applications.
\newblock \emph{Computer Methods in Applied Mechanics and Engineering},
  139\penalty0 (1--4):\penalty0 289 -- 314, 1996.
\newblock ISSN 0045-7825.
\newblock \doi{http://dx.doi.org/10.1016/S0045-7825(96)01087-0}.

\bibitem[Melenk and Wohlmuth(2001)]{Melenk2001}
J.~M. Melenk and B.~I. Wohlmuth.
\newblock On residual-based a posteriori error estimation in hp-fem.
\newblock \emph{Advances in Computational Mathematics}, 15\penalty0
  (1-4):\penalty0 311--331, 2001.
\newblock ISSN 1019-7168.
\newblock \doi{10.1023/A:1014268310921}.

\bibitem[Ming and Shi(2002)]{Ming}
P.~Ming and Z.-C. Shi.
\newblock Quadrilateral mesh revisited.
\newblock \emph{Computer methods in applied mechanics and engineering},
  191\penalty0 (49):\penalty0 5671--5682, 2002.

\bibitem[Mitchell and McClain(2014)]{Mitchell2014}
W.~F. Mitchell and M.~A. McClain.
\newblock A comparison of hp-adaptive strategies for elliptic partial
  differential equations.
\newblock \emph{ACM Transactions on Mathematical Software}, 41\penalty0
  (1):\penalty0 2, 2014.

\bibitem[Motamarri et~al.(2012)Motamarri, Nowak, Leiter, Knap, and
  Gavini]{Motamarri:2012wm}
P.~Motamarri, M.~R. Nowak, K.~Leiter, J.~Knap, and V.~Gavini.
\newblock {Higher-order adaptive finite-element methods for Kohn-Sham density
  functional theory}.
\newblock \emph{Journal of Computational Physics}, 253\penalty0 (15):\penalty0
  308--343, June 2012.

\bibitem[Mousavi et~al.(2012)Mousavi, Pask, and Sukumar]{Mousavi2012}
S.~Mousavi, J.~Pask, and N.~Sukumar.
\newblock Efficient adaptive integration of functions with sharp gradients and
  cusps in n-dimensional parallelepipeds.
\newblock \emph{International Journal for Numerical Methods in Engineering},
  91\penalty0 (4):\penalty0 343--357, 2012.

\bibitem[Pask et~al.(2011)Pask, Sukumar, Guney, and Hu]{Pask2011}
J.~Pask, N.~Sukumar, M.~Guney, and W.~Hu.
\newblock Partition-of-unity finite-element method for large scale quantum
  molecular dynamics on massively parallel computational platforms.
\newblock Technical report, Technical Report LLNL-TR-470692, Department of
  Energy LDRD 08-ERD-052, 2011.

\bibitem[Pask and Sterne(2005)]{Pask:2005bf}
J.~E. Pask and P.~A. Sterne.
\newblock {Finite element methods in ab initio electronic structure
  calculations}.
\newblock \emph{Modelling and Simulation in Materials Science and Engineering},
  13\penalty0 (3):\penalty0 R71--R96, Apr. 2005.
\newblock \doi{10.1088/0965-0393/13/3/R01}.

\bibitem[Patz{\'a}k and Jir{\'a}sek(2003)]{Patzak2003}
B.~Patz{\'a}k and M.~Jir{\'a}sek.
\newblock Process zone resolution by extended finite elements.
\newblock \emph{Engineering Fracture Mechanics}, 70\penalty0 (7):\penalty0
  957--977, 2003.

\bibitem[Simone(2007)]{Simone2007}
A.~Simone.
\newblock Partition of unity-based discontinuous finite elements: Gfem, pufem,
  xfem.
\newblock \emph{Revue Europ{\'e}enne de G{\'e}nie Civil}, 11\penalty0
  (7-8):\penalty0 1045--1068, 2007.

\bibitem[Sukumar and Pask(2009)]{Sukumar:2009vw}
N.~Sukumar and J.~E. Pask.
\newblock {Classical and enriched finite element formulations for
  Bloch-periodic boundary conditions}.
\newblock \emph{International Journal for Numerical Methods in Engineering},
  77\penalty0 (8):\penalty0 1121--1138, 2009.

\bibitem[Telles(1987)]{Telles1987}
J.~Telles.
\newblock A self-adaptive co-ordinate transformation for efficient numerical
  evaluation of general boundary element integrals.
\newblock \emph{International Journal for Numerical Methods in Engineering},
  24\penalty0 (5):\penalty0 959--973, 1987.

\bibitem[Veeser and Verf{\"u}rth(2011)]{Veeser}
A.~Veeser and R.~Verf{\"u}rth.
\newblock Poincar{\'e} constants for finite element stars.
\newblock \emph{IMA Journal of Numerical Analysis}, page drr011, 2011.

\bibitem[Verf{\"u}rth(1996)]{Verfuerth1996}
R.~Verf{\"u}rth.
\newblock A review of a posteriori error estimation and adaptive
  mesh-refinement techniques. 1996.
\newblock \emph{Teubner-Wiley, New York}, 1996.

\bibitem[Verf{\"u}rth(1999)]{Verf}
R.~Verf{\"u}rth.
\newblock Error estimates for some quasi-interpolation operators.
\newblock \emph{ESAIM: Mathematical Modelling and Numerical Analysis},
  33\penalty0 (04):\penalty0 695--713, 1999.

\bibitem[White et~al.(1989)White, Wilkins, and Teter]{White1989}
S.~R. White, J.~W. Wilkins, and M.~P. Teter.
\newblock Finite-element method for electronic structure.
\newblock \emph{Phys. Rev. B}, 39:\penalty0 5819--5833, Mar 1989.
\newblock \doi{10.1103/PhysRevB.39.5819}.

\bibitem[Xiao and Karihaloo(2006)]{Xiao2006}
Q.~Xiao and B.~Karihaloo.
\newblock Improving the accuracy of xfem crack tip fields using higher order
  quadrature and statically admissible stress recovery.
\newblock \emph{International Journal for Numerical Methods in Engineering},
  66\penalty0 (9):\penalty0 1378--1410, 2006.

\bibitem[Zhang et~al.(2008)Zhang, Shen, Zhou, and Gong]{Zhang:2008dp}
D.~Zhang, L.~Shen, A.~Zhou, and X.-G. Gong.
\newblock {Finite element method for solving Kohn{\textendash}Sham equations
  based on self-adaptive tetrahedral mesh}.
\newblock \emph{Physics Letters A}, 372\penalty0 (30):\penalty0 5071--5076,
  July 2008.
\newblock \doi{10.1016/j.physleta.2008.05.075}.

\end{thebibliography}

\end{document}